\documentclass[preprint,12pt]{elsarticle}
\usepackage{array}
\usepackage{amsmath}
\usepackage{graphics}
\usepackage{epsfig,amssymb,latexsym,verbatim}
\usepackage{graphicx}
\usepackage{multirow}
\usepackage{graphics}
\usepackage{amssymb}
\usepackage{amsthm}

\newtheorem{theorem}{Theorem}
\newtheorem{lemma}{Lemma}

\journal{Journal of Multivariate Analysis}

\def\T{{\mbox{\rm\tiny T}}}

\def\SBLL{{\mbox{\rm \tiny SBLL}}}
\def\LS{{\mbox{\rm \tiny LS}}}

\begin{document}

\begin{frontmatter}



\title{Nonparametric additive model-assisted estimation for survey data}

\author{Li Wang\corref{cor1}}
\ead{lilywang@uga.edu}
\cortext[cor1]{corresponding author}
\address{Department of Statistics, University of Georgia, Athens, Georgia 30602}

\author{Suojin Wang}
\address{Department of Statistics, Texas A\&M University, College Station,
Texas 77843}

\begin{abstract}
An additive model-assisted nonparametric method is investigated to
estimate the finite population totals of massive survey data with
the aid of auxiliary information. A class of estimators is
proposed to improve the precision of the well known
Horvitz-Thompson estimators by combining the spline and local
polynomial smoothing methods. These estimators are calibrated,
asymptotically design-unbiased, consistent, normal and robust in
the sense of asymptotically attaining the Godambe-Joshi lower
bound to the anticipated variance. A consistent model selection
procedure is further developed to select the significant auxiliary
variables. The proposed method is sufficiently fast to analyze
large survey data of high dimension within seconds. The
performance of the proposed method is assessed empirically via
simulation studies.

\end{abstract}


\end{frontmatter}


%
%
%
%

\section{Introduction}
\label{SEC:intro}

Auxiliary information is often available in many surveys for all
elements of the population of interest. For instance, in many
countries, administrative registers provide extensive sources of
auxiliary information. Complete registers can give access to
variables such as sex, age, income and country of birth. Studies
of labor force characteristics or household expenditure patterns,
for example, might benefit from these auxiliary data. Another
example is the satellite images or GPS data used in spatial
sampling. These data are often collected at the population level,
which are often available at little or no extra cost, especially
compared to the cost of collecting the survey data.

If no information other than the inclusion probabilities is used
to estimate the population total, a well-known design unbiased
estimator is the Horvitz-Thompson (HT) estimator. Nowadays,
``cheap'' auxiliary information can be regularly used to obtain
higher precision estimates for the unknown finite population
quantities. For instance, post-stratification, calibration and
regression estimation are different design-based approaches used
to improve the precision of estimators. Auxiliary information can
also be used to increase the accuracy of the finite population
distribution function; see, for example, \citep{WD97}.
Model-assisted estimation (\citep{SSW92}) provides a convenient
way to incorporate auxiliary variables to develop more efficient
survey estimators. By model-assisted, it is meant that a
superpopulation model is adopted (for example, model (\ref{model})
below), in which the finite population is modeled conditionally on
the auxiliary information; see, for instance,
\citep{C96,CDW93,D92,DH93}.

The traditional parametric model-assisted approach assumes that
the superpopulation model is fully described by a finite set of
parameters, e.g., the regression estimator introduced in \citep{SSW92}.
However, survey data now being
collected by many government, health and social science
organizations have more complex design features. It is difficult
to obtain any prior model information to address various
hypotheses. In this sense, preselected parametric model is too
restricted to fit unexpected features. In contrast, nonparametric
regression provides a useful tool for studying the dependence of
variables of interest on auxiliary information without
constraining the dependence to a fixed form with few parameters.
The flexibility of nonparametric regression is extremely helpful
to capture the complicated relationship between variables as well
as in obtaining robust predictions; see \citep{FG96,H90} for details.

Breidt and Opsomer \citep{BO00} first proposed a nonparametric
model-assisted estimator based on local polynomial regression,
which generalizes the parametric framework in survey sampling and
improves the precision of the survey estimators immensely. Their
investigation is only based on one auxiliary variable. Most
surveys, however, involve more than one study variables, perhaps,
many; see \citep{SL05}. For example, the
remote sensing data which provide a wide and growing range of
variables to be employed. In this context, when the dimension of
the auxiliary information vector is high, one unavoidable issue is
the ``curse of dimensionality'', which refers to the poor
convergence rate of nonparametric estimation of general
multivariate functions. One solution is regression in the form of
additive model; see \citep{HT90}.

Estimation and inference for additive models have been well
studied in the literature; see, for example, the classic
backfitting estimators of \citep{HT90}, the
marginal integration estimators of \citep{LN95}, the
smoothing backfitting estimators of \citep{MLN99}, the
spline estimators of Stone (\citep{S85,S94}) and the spline-backfitted
kernel estimators of \citep{WY07}. In survey sampling
context, \citep{BCO05} discussed a semiparametric possible
extension to multiple auxiliary variables via using the penalized
splines; \citep{OBMK07} applied the generalized additive
models (GAMs) in an interaction model for the estimation of
variables from forest inventory and analysis surveys; and \citep{BOJR07}
proposed a special case of the GAMs with an identity
link function. For large and high dimensional survey data, it is
important that estimation and inference methods are efficient and
computationally easily implemented. However, few methods are
theoretically justified and computational efficient when there are
multiple nonparametric terms. The kernel based backfitting and
marginal integration approaches are computationally expensive,
limiting their use for high dimensional data; see \citep{MFY07}
for some numerical comparisons of these methods.
Spline methods, on the other hand, provide only convergence rates
but no asymptotic distributions, so no measures of confidence can
be assigned to the estimators.

Challenged by these demands, we propose approximating the
nonparametric components by using the spline-backfitted local
polynomial: spline does a quick initial estimation of all additive
components and removes them all except the ones of interest;
kernel smoothing is then applied to the cleaned univariate data to
estimate with the asymptotic distribution. This two-step estimator
is both computationally expedient for analyzing large and high
dimensional survey data, and theoretically reliable as the
estimator is uniformly oracle with asymptotic confidence
intervals. The resulting estimator of population total can
therefore be easily calculated, and more importantly allow for
formal derivation of the asymptotic properties of the estimator.

In practice, a large number of variables may be collected and some
of the insignificant ones should be excluded from the final model
in order to enhance the predictability. The selection of auxiliary
variables is a fundamental issue for model-assisted survey
sampling methods. In this paper, we propose a consistent variable
selection method for the additive model-assisted survey sampling
based on the Bayes information criterion (BIC). A comprehensive
Monte Carlo study demonstrates superior performance of the
proposed methods.

The rest of the paper is organized as follows. Section 2 gives
details of the superpopulation model and proposed method of
estimation. Section 3 describes the weighting, calibration and
asymptotic properties of the proposed estimator. Section 4
describes the auxiliary variable selection procedure for the
superpopulation model under simple random sampling design (SRS).
Section 5 reports the findings in an extensive simulation study.
Lengthy technical arguments are given in the Appendix.

\section{Superpopulation Model and Proposed Estimator}
\label{SEC:ModelEstimation}

In what follows, let $U_{N}=\left\{ 1,...,i,...,N\right\}$ be the
finite population of $N$ elements, called the target population,
and $i$ represents the $i$th element of the population.
Let $\mathbf{x}_{i}=\left\{ x_{i1},...,x_{id}\right\} $ be a
$d$-dimensional auxiliary variable vector, $i\in U_{N}$. We are
interested in the estimation of the population total
$t_{y}=\sum_{i\in U_{N}}y_{i}$, where $y_{i}$ is the value of the
study variable, $y$, for the $i$th element.  To this end, a sample
$s$ of size $n_{N}$ is drawn from $U_{N}$
according to a fixed sampling design $p_{N}\left( \cdot \right) $, where $%
p_{N}\left( s\right) $ is the probability of drawing the sample
$s$. The inclusion probabilities, known for all $i\in U_{N}$, are
$\pi _{iN}\equiv \pi _{i}=\Pr \left\{ i\in s\right\} =\sum_{s\ni
i}p_{N}\left( s\right) $. In addition to the $\pi _{i}$, denote
$\pi _{ijN}\equiv \pi _{ij}=\Pr \left\{ i,j\in s\right\} =\sum_{s
\ni i,j}p_{N}\left( s\right) $ the inclusion probability for both
elements $i,j\in U_{N}$.

Let $\left\{ \left( \mathbf{x}_{i},y_{i}\right) \right\} _{i\in
U_{N}}$ be a realization of $\left( \mathbf{X},Y\right) $ from an
infinite superpopulation, $\xi $, satisfying
\begin{equation}
Y=m\left( \mathbf{X}\right) +\sigma \left( \mathbf{X}\right)
\varepsilon , \label{model}
\end{equation}
in which the unknown $d$-variate function $m$ has a simpler form
of
\begin{equation}
m\left( \mathbf{X}\right) =c+\sum_{\alpha =1}^{d}m_{\alpha }\left(
X_{\alpha }\right),\mbox{\quad} E_{\xi}\left[m_{\alpha
}\left(X_{\alpha }\right)\right] \equiv 0,\mbox{\quad} 1\leq
\alpha \leq d, \label{model:am}
\end{equation}
the function $\sigma(\cdot)$ is the unknown standard deviation
function and the standard error $\varepsilon$ satisfies that
$E_{\xi }\left( \varepsilon \left\vert \mathbf{X}\right. \right)
=0$ and $E_{\xi }\left( \varepsilon ^{2}\left\vert
\mathbf{X}\right. \right) =1$. In the following, we assume the
auxiliary variable $X_{\alpha }$ is distributed on a compact
interval $\left[ a_{\alpha },b_{\alpha }\right]$, $\alpha
=1,...,d$. Without loss of generality, we take all intervals
$\left[ a_{\alpha },b_{\alpha }\right] =\left[ 0,1\right]$. To
estimate the additive components in (\ref{model:am}), we employ a
two-stage procedure based on the spline-backfitted local
polynomial smoothing.

For any $\alpha =1,...,d$, we introduce a knot sequence with $J$
interior knots $k_{0\alpha }=0<k_{1\alpha }<...<k_{J\alpha }<1=k_{\left(
J+1\right) \alpha }$, where $J\equiv J_{N}$ increases when $n_{N}$ increases, and the
precise order is given in Assumption (A5). Denote the piecewise
linear truncated power spline basis
\begin{equation}
\mathbf{\Gamma }\left(\mathbf{x}\right)\equiv \left\{ 1,x_{\alpha
},\left( x_{\alpha }-k_{1\alpha }\right) _{+},\ldots ,\left(
x_{\alpha }-k_{J\alpha }\right) _{+},\alpha
=1,...,d\right\}^{\T},\label{DEF:Gamma(x)}
\end{equation}
where $\left(a\right) _{+}=a$ if $a>0$ and $0$ otherwise. For the
local linear smoothing, let $K_{h}\left( x\right)
=h^{-1}K\left(x/h\right) $, where $K$ denotes a kernel function
and $h=h_{N}$ is the bandwidth; see Assumption (A6) below.

We now describe our two-stage estimator for the population total $t_{y}$. At
the first stage, we apply the spline smoothing to obtain a quick initial
estimator of $m\left( \mathbf{x}_{i}\right)$,
\[
\hat{m}\left( \mathbf{x}_{i}\right) =\hat{b}_{0}+\sum_{\alpha =1}^{d}%
\hat{b}_{0,\alpha }x_{i\alpha }+\sum_{\alpha =1}^{d}\sum_{j=1}^{J}\hat{b}%
_{j,\alpha }\left( x_{i\alpha }-k_{j\alpha }\right) _{+},
\]
where $\hat{b}_{0}$ and $\hat{b} _{j,\alpha}$, $j=0,1,..,J$, $\alpha=1,...,d$
are the minimizes of the following
\begin{equation}
\sum_{i\in s}\pi_{i}^{-1}\left\{ y_{i}-b_{0}-\sum_{\alpha
=1}^{d}b_{0,\alpha }x_{\alpha }-\sum_{\alpha
=1}^{d}\sum_{j=1}^{J}b_{j,\alpha }\left( x_{i\alpha }-k_{j\alpha
}\right) _{+}\right\} ^{2} \label{EQ:least square 1}
\end{equation}
over a $G_{d}\equiv 1+(J+1)d$ dimensional vector. Because the components
$m_{\alpha }\left( x_{\alpha }\right)$ can only be identified up to an additive
constants, we center the estimator of $m_{\alpha }\left( x_{\alpha }\right)$
and define the centered pilot estimator of the $\alpha$th component as
\begin{equation}
\hat{m}_{\alpha }\left( x_{\alpha }\right) =\hat{b}%
_{0,\alpha }x_{\alpha }+\sum_{j=1}^{J}\hat{b}_{j,\alpha }\left(
x_{\alpha }-k_{j\alpha }\right) _{+}-\hat{c}_{\alpha },
\label{DEF:mhatalpha-xalpha}
\end{equation}
where $ \hat{c}_{\alpha }=N^{-1}\sum_{i\in s}
\pi_{i}^{-1}\left\{\hat{b}_{0,\alpha }x_{i\alpha
}+\sum_{j=1}^{J}\hat{b}_{j,\alpha }\left( x_{i\alpha }-k_{j\alpha
}\right) _{+}\right\}$. The above pilot estimators in
(\ref{DEF:mhatalpha-xalpha}) are then used to construct the new
pseudo-responses
\begin{equation}
\hat{y}_{i\alpha }=y_{i}-N^{-1}\hat{t}_{y}-\sum_{\beta \neq \alpha
} \hat{m}_{\beta}(x_{i\alpha}),\ i \in s, \ \alpha=1,...,d,
\label{DEF:yhat}
\end{equation}
where $\hat{t}_{y}$ is the well-known HT estimator.

At the second stage, a local polynomial smoothing is applied to the cleaned
univariate data $\left\{ x_{i\alpha },\hat{y}_{i\alpha }\right\} _{i\in s}$ to
achieve the ``oracle'' property in \citep{WY07}. To be specific, considering
the local linear smoothing, for any $\alpha=1$, ..., $d$, we minimize
\begin{equation}
\sum_{i\in s}\pi_{i}^{-1}\left\{
\hat{y}_{i\alpha}-a_{0,\alpha}-a_{1,\alpha}\left( x_{i\alpha
}-x\right)K_{h}\left(x_{i\alpha}-x\right)\right\}^2,
\label{EQ:least square 2}
\end{equation}
with respect to $a_{0,\alpha}$ and $a_{1,\alpha}$. The spline-backfitted local
linear (SBLL) estimator of the $\alpha$-th component $m_{\alpha}$ is
$\hat{m}_{\alpha }^{*}=\hat{a}_{0,\alpha}$ in (\ref{EQ:least square 2}). The
final sample design-based SBLL estimator of ${m}\left(\mathbf{x}\right)$ is
defined as
\begin{equation}
\hat{m}^{*}\left(\mathbf{x}\right)=\frac{1}{N}\hat{t}_{y}+\sum_{\alpha
=1}^{d}\hat{m}_{\alpha }^{*}\left(x_\alpha\right).
\label{DEF:mihatstar}
\end{equation}

Substituting $\hat{m}_{i}^{*}\equiv
\hat{m}^{*}\left(\mathbf{x}_{i}\right)$ into the existing
generalized difference estimator (see page 221 of \citep{SSW92}),
the SBLL estimator for $t_{y}$ is defined by
\begin{equation}
\hat{t}_{y,\SBLL}=\sum_{i\in U_{N}}\hat{m}_{i}^{*}+
\sum_{i\in s}\frac{y_{i}-\hat{m}_{i}^{*}}{\pi _{i}}%
= \sum_{i\in s}\frac{y_{i}}{\pi _{i}}%
+\sum_{i\in
U_{N}}\left(1-\frac{I_{i}}{\pi_{i}}\right)\hat{m}_{i}^{*},
\label{DEF:tySBLLhat}
\end{equation}
where $I_{i}=1$ if $i\in s$ and $I_{i}=0$ otherwise.

\vskip .1in \noindent\textbf{Remark 1.} In the first step spline
smoothing, the number of knots $J_{N}$ can be determined by
$n_{N}$ and a tuning constant $c$:
\begin{equation}
J_{N}=\min \left( [cn_{N}^{1/4}\log (n_{N})] +1,\left[
\left( n_{N}/2-1\right) /d-1\right] \right).
\label{DEF:Numberofknots}
\end{equation}
As discussed in \citep{WY07}, the choice of $c$ makes little
difference. In the second step local polynomial smoothing, one can
use the quartic kernel and the rule-of-thumb bandwidth.

\section{Properties of the Estimator}
\label{SEC:Property}

\subsection{Weighting and Calibration}

In the last decade, calibration estimation has developed into an important
field of research in survey sampling. As discussed in \citep{DS92} and
\citep{K09}, calibration is a highly desirable property for survey weights,
which allows the survey practitioner to simply adjust the original design
weights to incorporate the information of the auxiliary variables. Several
national statistical agencies have developed software to compute calibrated
weights based on auxiliary information available in population registers and
other sources. The proposed SBLL estimator in this paper also shares this
property in certain sense.

Let $\mathbf{y}_{s}$ be the column vector of the response values $y_{i}$ for
$i\in s$ and define the diagonal matrix of inverse inclusion probabilities
$\mathbf{\Pi }_{s}=\text{diag}\left\{ 1/{\pi _{i}}\right\} _{i\in s}$. For
$\mathbf{\Gamma}(\mathbf{x})$ in (\ref{DEF:Gamma(x)}), denote $\mathbf{\Gamma}
_{s}=\left\{ \mathbf{\Gamma }\left(\mathbf{x}_{i} \right)^{\T}\right\}_{i\in
s}$ the sample truncated power spline matrix. Let $\mathbf{B}_{s}$ be the
collection of the estimated spline coefficient in (\ref{EQ:least square 1}),
then
$\mathbf{B}_{s}=\left( \mathbf{\Gamma }_{s}^{\T}\mathbf{\Pi }_{s}\mathbf{%
\Gamma }_{s}\right) ^{-1}\mathbf{\Gamma }_{s}^{\T}\mathbf{\Pi }_{s}\mathbf{y}%
_{s}$. Thus the pilot spline estimator of $m_{\alpha }\left(
x_{\alpha }\right)$ in (\ref{DEF:mhatalpha-xalpha}) can be written
as
\begin{equation}
\hat{m}_{\alpha }\left( x_{\alpha }\right) =\left\{\mathbf{\Gamma
}\left(\mathbf{x}\right)^{\T}\mathbf{D}_{\alpha}\mathbf{B}_{s}-N^{-1}\mathbf{1}_{n}^{\T}%
\mathbf{\Pi }_{s} \mathbf{\Gamma }_{s}\mathbf{D}_{\alpha }\mathbf{B}%
_{s}\right\}\mathbf{y}_{s}, \label{EQ:mhatalpha-xalpha}
\end{equation}
where $\mathbf{1}_{n}$ is a vector of length $n_{N}$ with all
``1''s, and
\begin{equation}
\mathbf{D}_{\alpha }=\text{diag}\{ 0,...,0,\mathrel{\mathop{%
\underbrace{1,...,1},}\limits_{\text{from }\left( J+1\right)\left(
\alpha -1\right) +2\text{ to }\left(J+1\right)\alpha +1}}0,...,0\}
\label{DEF:Dalpha}
\end{equation}
is a $G_{d}\times G_{d}$ diagonal matrix. Denoting the spline
smoothing matrix and its centered version by
\[
\mathbf{\Psi }_{s\alpha }=\mathbf{\Gamma }_{s}\mathbf{D}_{\alpha
}\left(
\mathbf{\Gamma }_{s}^{\T}\mathbf{\Pi }_{s}\mathbf{\Gamma }_{s}\right) ^{-1}%
\mathbf{\Gamma }_{s}^{\T}\mathbf{\Pi }_{s}, \mbox{\quad}
\mathbf{\Psi }_{s\alpha }^{*}=\left( \mathbf{I}-N^{-1}\mathbf{1}_{n}\mathbf{1}_{n}^{\T}%
\mathbf{\Pi }_{s}\right) \mathbf{\Psi }_{s\alpha },
\]
we have $\hat{\mathbf{m}}_{\alpha }\equiv \left\{ \hat{m}_{\alpha }\left(
x_{i\alpha }\right) \right\} _{i\in s} =\mathbf{\Psi }_{s\alpha
}^{*}\mathbf{y}_{s}$, for $\alpha =1,...,d$. Further for $\hat{y}_{i\alpha }$
in (\ref{DEF:yhat}), let $\hat{\mathbf{y}}_{\alpha }\equiv \left\{
\hat{y}_{i\alpha }\right\} _{i\in
s}=\mathbf{y}_{s}-\frac{1}{N}\hat{t}_{y}\mathbf{1}_{n}-\sum_{\beta \neq \alpha
}\hat{\mathbf{m}}_{\beta }$, and define the matrices
\[
\mathbf{X}_{si\alpha }=\left\{\left(
\begin{array}{ll}
1 & x_{k\alpha }-x_{i\alpha }
\end{array}\right)
\right\} _{k\in s}, \mbox{\quad} \mathbf{W}_{si\alpha }=\text{diag}\left\{ \frac{1}{\pi _{k}%
}K_{h}\left( x_{k\alpha }-x_{i\alpha }\right) \right\} _{k\in s}.
\]
Then the SBLL estimator of $m_{\alpha}$ at $x_{i\alpha}$ can be written as
\begin{equation}
\hat{m}_{i\alpha }^{*}\equiv \hat{m}_{\alpha
}^{*}\left(x_{i\alpha}\right)=\mathbf{e}_{1}^{\T}\left(
\mathbf{X}_{si\alpha
}^{\T}\mathbf{W}_{si\alpha }\mathbf{X}_{si\alpha }\right) ^{-1}\mathbf{X}%
_{si\alpha }^{\T}\mathbf{W}_{si\alpha }\hat{\mathbf{y}}_{\alpha },
\label{DEF:mhatstar-ialpha}
\end{equation}
where $\mathbf{e}_{1}=\left(1,0\right)^{\T}$. Therefore, the SBLL estimator in
(\ref{DEF:mihatstar}) of ${m}\left(\mathbf{x}\right)$ at $\mathbf{x}_{i}$ is
\begin{eqnarray*}
\hat{m}_{i}^{*}&=&\frac{1}{N}\hat{t}_{y}+\sum_{\alpha =1}^{d}\mathbf{e}%
_{1}^{\T}\left( \mathbf{X}_{si\alpha }^{\T}\mathbf{W}_{si\alpha }\mathbf{X}%
_{si\alpha }\right) ^{-1}\mathbf{X}_{si\alpha
}^{\T}\mathbf{W}_{si\alpha }\left(
\mathbf{y}_{s}-\frac{\hat{t}_{y}}{N}\mathbf{1}_{n}-\sum_{\beta
\neq \alpha }\mathbf{\Psi }_{s\beta }^{*}\mathbf{y}_{s}\right)\\
 &\equiv& \mathbf{\rho
}_{si}^{\T}\mathbf{y}_{s},
\end{eqnarray*}
where\\ $\mathbf{\rho }_{si}^{\T}=\mathbf{e}_{1}^{\T}\left\{ \sum_{\alpha
=1}^{d}\left( \mathbf{X}_{si\alpha }^{\T}\mathbf{W}_{si\alpha
}\mathbf{X}_{si\alpha
}\right) ^{-1}\mathbf{X}_{si\alpha }^{\T}\mathbf{W}_{si\alpha }\left( \mathbf{%
I}+\frac{1-d}{dN}\mathbf{1}_{n}\mathbf{1}_{n}^{\T}%
\mathbf{\Pi }_{s}-\sum_{\beta \neq \alpha
}\mathbf{\Psi }_{s\beta }^{*}\right) \right\}$.

Similar to \citep{SSW89}, we define the ``g-weight''
\begin{equation}
g_{is}=1+\pi
_{i}\sum_{j\in U_{N}}\left( 1-\frac{I_{j}}{\pi _{j}}\right) \mathbf{\rho }%
_{sj}^{{\mbox{\rm\tiny T}}}\mathbf{a}_{i},
\label{DEF:gweight}
\end{equation}
where $\mathbf{a}_i$ is a
$n_{N}$-dimensional vector with a ``1'' in the $i$th position and ``0''
elsewhere.  Thus the proposed estimator $\hat{t}_{y,{\mbox{\rm \tiny SBLL}}}$
in (\ref{DEF:mihatstar}) can be written as
\[
\hat{t}_{y,{\mbox{\rm \tiny SBLL}}} = \sum_{i\in s}\frac{y_{i}}{\pi _{i}}%
+\sum_{j\in U_{N}}\left( 1-\frac{I_{j}}{\pi _{j}}\right) \mathbf{\rho }%
_{sj}^{{\mbox{\rm\tiny T}}}\mathbf{y}_{s}\equiv \sum_{i\in s}\left. g_{is}y_{i}\right/ \pi _{i},
\]
which is a linear combination of the sample $y_{i}$'s with a sampling weight,
$\pi _{i}^{-1}$, and the ``g-weight''. Because the weights are independent of
$y_{i}$, they can be applied to any study variable of interest.

As we show below, the weight system gives our estimator of the known total
$\sum_{i\in U_{N}}x_{i\alpha }$ to be itself.

\begin{theorem}
For any $\alpha =1,...,d$ and the ``g-weight'' defined in (\ref{DEF:gweight}),
\[
\hat{t}_{x_{\alpha },{\mbox{\rm \tiny SBLL}}}\equiv \sum_{i\in
s}g_{is}x_{i\alpha }/\pi _{i}=\sum_{i\in U_{N}}x_{i\alpha }.
\]
\end{theorem}

\noindent \textbf{Proof.} Let $\mathbf{x}_{\alpha }=\left\{ x_{i\alpha
}\right\} _{i\in s}$. We have
\begin{eqnarray*}
&&\hat{t}_{x_{\alpha },{\mbox{\rm \tiny SBLL}}}=\sum_{i\in s}\pi
_{i}^{-1}x_{i\alpha }+\sum_{j\in U_{N}}\left( 1-I_{j}\pi _{j}^{-1}%
\right) \mathbf{e}_{1}^{{\mbox{\rm\tiny T}}} \\
&&\times \left\{ \sum_{\gamma =1}^{d}\left( \mathbf{X}_{sj\gamma }^{{%
\mbox{\rm\tiny T}}}\mathbf{W}_{sj\gamma }\mathbf{X}_{sj\gamma }\right) ^{-1}%
\mathbf{X}_{sj\gamma }^{{\mbox{\rm\tiny T}}}\mathbf{W}_{sj\gamma }\left(
\mathbf{I}+\frac{1-d}{dN}\mathbf{1}_{n}\mathbf{1}_{n}^{{\mbox{\rm\tiny T}}}%
\mathbf{\Pi }_{s}-\sum_{\beta \neq \gamma }\mathbf{\Psi }_{s\beta
}^{*}\right) \right\} \mathbf{x}_{\alpha }.
\end{eqnarray*}
Observe that
\[
\left( \mathbf{\Gamma }_{s}^{{\mbox{\rm\tiny T}}}\mathbf{\Pi }_{s}\mathbf{%
\Gamma }_{s}\right) ^{-1}\mathbf{\Gamma }_{s}^{{\mbox{\rm\tiny T}}}\mathbf{%
\Pi }_{s}\mathbf{x}_{\alpha }={\mbox{\boldmath $\tau$}}_{\alpha },%
\mbox{\quad}\mathbf{\Psi }_{s\beta }\mathbf{x}_{\alpha }=\mathbf{\Gamma }_{s}%
\mathbf{D}_{\beta }{\mbox{\boldmath $\tau$}}_{\alpha }=\left\{
\begin{array}{ll}
\mathbf{x}_{\alpha }, & \text{for }\beta =\alpha  \\
\mathbf{0}, & \text{for }\beta \neq \alpha
\end{array}
,\right.
\]
where ${\mbox{\boldmath $\tau$}}_{\alpha }$ is the vector of dimension $G_{d} $
with a ``1'' in the $\left\{ 2+(J+1)\left( \alpha -1\right) \right\} $th
position and ``0'' elsewhere. Then we have,
\[
\left( \mathbf{I}+\frac{1-d}{dN}\mathbf{1}_{n}\mathbf{1}_{n}^{{%
\mbox{\rm\tiny T}}}\mathbf{\Pi }_{s}-\sum_{\beta \neq \gamma }\mathbf{\Psi }%
_{s\beta }^{*}\right) \mathbf{x}_{\alpha }=\left\{
\begin{array}{lc}
\left( \mathbf{I}+\frac{1-d}{dN}\mathbf{1}_{n}\mathbf{1}_{n}^{{%
\mbox{\rm\tiny T}}}\mathbf{\Pi }_{s}\right) \mathbf{x}_{\alpha }, & \text{%
for }\gamma =\alpha  \\
\frac{1}{dN}\mathbf{1}_{n}\mathbf{1}_{n}^{{\mbox{\rm\tiny T}}}\mathbf{\Pi }%
_{s}\mathbf{x}_{\alpha }, & \text{for }\gamma \neq \alpha
\end{array}
.\right.
\]
Note that for any $i\in U_{N}$,
\[
\mathbf{e}_{1}^{{\mbox{\rm\tiny T}}}\left( \mathbf{X}_{si\alpha }^{{%
\mbox{\rm\tiny T}}}\mathbf{W}_{si\alpha }\mathbf{X}_{si\alpha }\right) ^{-1}%
\mathbf{X}_{si\alpha }^{{\mbox{\rm\tiny T}}}\mathbf{W}_{si\alpha }\mathbf{x}%
_{\alpha } =x_{i\alpha }, \
\mathbf{e}_{1}^{{\mbox{\rm\tiny T}}}\left( \mathbf{X}_{si\alpha }^{{%
\mbox{\rm\tiny T}}}\mathbf{W}_{si\alpha }\mathbf{X}_{si\alpha }\right) ^{-1}%
\mathbf{X}_{si\alpha }^{{\mbox{\rm\tiny T}}}\mathbf{W}_{si\alpha }\mathbf{1}%
_{n} =1,
\]
thus
\begin{eqnarray*}
&&\mathbf{e}_{1}^{{\mbox{\rm\tiny T}}}\left\{ \sum_{\gamma =1}^{d}\left(
\mathbf{X}_{sj\gamma }^{{\mbox{\rm\tiny
T}}}\mathbf{W}_{sj\gamma }\mathbf{X}_{sj\gamma }\right) ^{-1}\mathbf{X}%
_{sj\gamma }^{{\mbox{\rm\tiny T}}}\mathbf{W}_{sj\gamma }\left( \mathbf{I}+%
\frac{1-d}{dN}\mathbf{1}_{n}\mathbf{1}_{n}^{{\mbox{\rm\tiny T}}}\mathbf{\Pi }%
_{s}-\sum_{\beta \neq \gamma }\mathbf{\Psi }_{s\beta }^{*}\right) \right\}
\mathbf{x}_{\alpha } \\
&=&\mathbf{e}_{1}^{{\mbox{\rm\tiny T}}}\left( \mathbf{X}_{si\alpha }^{{%
\mbox{\rm\tiny T}}}\mathbf{W}_{si\alpha }\mathbf{X}_{si\alpha }\right) ^{-1}%
\mathbf{X}_{si\alpha }^{{\mbox{\rm\tiny T}}}\mathbf{W}_{si\alpha }\left\{
\mathbf{I}+(dN)^{-1}(1-d)\mathbf{1}_{n}\mathbf{1}_{n}^{{\mbox{\rm\tiny T}}}%
\mathbf{\Pi }_{s}\right\} \mathbf{x}_{\alpha } \\
&&+{d}^{-1}\mathbf{e}_{1}^{{\mbox{\rm\tiny T}}}\sum_{\gamma \neq \alpha }\left(
\mathbf{X}_{sj\gamma }^{{\mbox{\rm\tiny T}}}\mathbf{W}_{sj\gamma }\mathbf{X}%
_{sj\gamma }\right) ^{-1}\mathbf{X}_{sj\gamma }^{{\mbox{\rm\tiny T}}}\mathbf{%
W}_{sj\gamma }\mathbf{1}_{n}\mathbf{1}_{n}^{{\mbox{\rm\tiny T}}}%
\mathbf{\Pi }_{s}\mathbf{x}_{\alpha } = x_{j\alpha }.
\end{eqnarray*}
Hence the proposed SBLL estimator defined in (\ref{DEF:tySBLLhat}) preserves
the calibration property. \hfill $\square$

\subsection{Assumptions}

For the asymptotic properties of the estimators, we adopt the
traditional asymptotic framework in \citep{BO00} where both the
population and sample sizes increase as $N\rightarrow \infty$.
There are two sources of ``variation'' to be considered here. The
first is introduced by the random sample design and the
corresponding measure is denoted by $p$. The ``$O_{p}$'',
``$o_{p}$'' and ``$E_{p}(\cdot)$'' notation below is with respect
to this measure. The second is associated with the superpopulation
from which the finite population is viewed as a sample. The
corresponding measure and notation are ``$\xi$''. For simplicity,
let $\pi _{ij}-\pi _{i}\pi _{j}=\Delta_{ij}$.

\begin{enumerate}
\item[(A1)]  \textit{The density }$f\left( \mathbf{x}\right)$%
\textit{\ of }$\mathbf{X}$\textit{\ is continuous and bounded away
from $0$ and $\infty$. The marginal densities }$f_{\alpha }\left( x_{\alpha }\right) $%
\textit{\ of }$x_{\alpha }$\textit{\ have continuous derivatives
and are bounded away from $0$ and $\infty$.}

\item[(A2)]  \textit{The second order derivative of $m_{\alpha
}\left( x_{\alpha }\right)$ is continuous, $\forall$ $\alpha
=1,...,d$.}

\item[(A3)]  \textit{There exists a positive constant} $M$
\textit{such that} $E_{\xi }\left( \left| \varepsilon \right|
^{2+\delta }\left| \mathbf{X}\right. \right) <M$\textit{\ for some
}$\delta
>1/2$; $\sigma \left( \mathbf{x}\right) $%
\textit{\ is continuous on }$\left[ 0,1\right] ^{d}$ \textit{and
bounded away from $0$ and $\infty$.}

\item[(A4)]  \textit{As} $N\rightarrow \infty $, $n_{N} \rightarrow \infty$ and
$n_{N}N^{-1}\rightarrow \pi < 1$.

\item[(A5)]  \textit{The number of knots} $J_{N}\thicksim
n_{N}^{1/4}\log (n_{N})$.

\item[(A6)]  \textit{The kernel function }$K$ \textit{is Lipschitz
continuous, bounded, nonnegative, symmetric, and supported on
}$\left[ -1,1\right] $.
\textit{The bandwidth} $h_{N}\thicksim n_N^{-1/5}$, \textit{i.e., }$%
c_{h}n_{N}^{-1/5}\leq h_{N}\leq C_{h}{n}_{N}^{-1/5}$\textit{\ for
some positive constants} $c_{h}$\textit{,} $C_{h}$.

\item[(A7)]  \textit{For all} $N$, $\min_{i\in U_{N}}\pi _{i}\geq
\lambda
>0, $ $\min_{i,j\in U_{N}}\pi _{ij}\geq \lambda ^{*}>0$ \textit{and}
\[
\limsup_{N\rightarrow \infty }n_{N}\max_{i,j\in U_{N},i\neq
j}\left| \Delta_{ij}\right| <\infty .
\]

\item[(A8)]  \textit{Let} $D_{k,N}$ \textit{be the set of all distinct} $k$%
\textit{-tuples} $\left( i_{1},i_{2},...,i_{k}\right) $
\textit{from} $U_{N}$. Then
\[
\hspace{-1cm}\limsup_{N\rightarrow \infty }n_{N}^{2}\max_{\left(
i_{1},i_{2},i_{3},i_{4}\right) \in D_{4,N}}\left| E_{p}\left[
\left( I_{i_{1}}-\pi _{i_{1}}\right) \left( I_{i_{2}}-\pi
_{i_{2}}\right) \left( I_{i_{3}}-\pi _{i_{3}}\right) \left(
I_{i_{4}}-\pi _{i_{4}}\right) \right] \right| <\infty ,
\]
\[
\limsup_{N\rightarrow \infty }n_{N}^{2}\max_{\left(
i_{1},i_{2},i_{3},i_{4}\right) \in D_{4,N}}\left| E_{p}\left[
\left( I_{i_{1}}I_{i_{2}}-\pi _{i_{1}i_{2}}\right) \left(
I_{i_{3}}I_{i_{4}}-\pi _{i_{3}i_{4}}\right) \right] \right|
<\infty ,
\]
\[
\limsup_{N\rightarrow \infty }n_{N}^{2}\max_{\left(
i_{1},i_{2},i_{3}\right) \in D_{3,N}}\left| E_{p}\left[ \left(
I_{i_{1}}-\pi _{i_{1}}\right) ^{2}\left( I_{i_{2}}-\pi
_{i_{2}}\right) \left( I_{i_{3}}-\pi _{i_{3}}\right) \right]
\right| <\infty .
\]
\end{enumerate}

\vskip .1in \noindent \textbf{Remark 2.} Assumptions (A1)-(A3) are typical in
the smoothing literature; see, for instance, \citep{FG96,H90,WY07}. Assumption
(A5) is about how to choose the number of interior knots $J_{N}$ for the spline
estimation in the first stage. In practice, $J_{N}$ can be determined by
(\ref{DEF:Numberofknots}). Assumption (A6) is how to select the kernel function
and the corresponding bandwidth. Such assumptions were used in \citep{WY07} in
the additive autoregressive model fitting. Assumptions (A7) and (A8) involve
the inclusion probabilities of the design, which were also assumed in
\citep{BO00}.

\subsection{Asymptotic properties of the estimator}

Like the local polynomial estimators in \citep{BO00},
the following theorem shows that the estimator $\hat{t}_{y,\SBLL}$
in (\ref{DEF:tySBLLhat}) is asymptotically design unbiased and
design consistent.

\begin{theorem}
\label{THM:ADU} Under Assumptions (A1)-(A7), the estimator
$\hat{t}_{y,\SBLL}$ in (\ref{DEF:tySBLLhat}) is asymptotically
design unbiased in the sense that
\[
\lim_{N\rightarrow \infty }E_{p}\left[ \frac{\hat{t}_{y,\SBLL}-t_{y}}{%
N}\right] =0\text{ with }\xi \text{-probability }1,
\]
and is design consistent in the sense that for all $\eta >0$,
\[
\lim_{N\rightarrow \infty }E_{p}\left[ I_{\left\{ \left|
\hat{t}_{y,\SBLL}-t_{y}\right| >N\eta \right\} }\right]=0\text{ with }\xi \text{
-probability }1.
\]
\end{theorem}

Let $\tilde{t}_{y,\SBLL}$ be the population-based generalized
difference estimator of $t_{y}$ when the entire realization were
known; see (\ref{DEF:tySBLLtilde}) in Appendix A.1 for the formal
definition. Like the estimators in the local polynomial estimators
in \citep{BO00}, the penalized spline estimators in
\citep{BCO05}, and the backfitting estimators in \citep{BOJR07},
the following theorem shows that the proposed
estimator $\hat{t}_{y,\SBLL}$ also inherits the limiting
distribution of the ``oracle'' estimator $\tilde{t}_{y,\SBLL}$.

\begin{theorem}
\label{THM:normality} Under Assumptions (A1)-(A8),
\[
\frac{N^{-1}\left( \tilde{t}_{y,\SBLL}-t_{y}\right) }{\mathrm{Var}%
_{p}^{1/2}\left( N^{-1}\tilde{t}_{y,\SBLL}\right) }\stackrel{d}{%
\longrightarrow }N\left( 0,1\right)
\]
as $N\rightarrow \infty $ implies
\[
\frac{N^{-1}\left( \hat{t}_{y,\SBLL}-t_{y}\right) }{\widehat{V}%
^{1/2}\left( N^{-1}\hat{t}_{y,\SBLL}\right) }\stackrel{d}{%
\longrightarrow }N\left( 0,1\right) ,
\]
where
\begin{equation}
\widehat{V}\left( N^{-1}\hat{t}_{y,\SBLL}\right) =\frac{1}{N^{2}}%
\sum_{i,j\in s}\frac{\Delta_{ij}}{\pi
_{ij}}\frac{y_{i}-\hat{m}_{i}^{*}}{\pi _{i}} \frac{ y_{j}-\hat{m}
_{j}^{*}}{\pi _{j}} . \label{DEF:Vhat}
\end{equation}
\end{theorem}

The next theorem proves that $\hat{t}_{y,\SBLL}$ is robust as in \citep{BO00} and
it also asymptotically attains the Godambe-Joshi lower bound to
the anticipated variance
\[
\text{Var}\left[N^{-1}\left(\hat{t}_{y}-t_{y}\right)\right]
=E\left[N^{-1}\left(\hat{t}_{y}-t_{y}\right)\right]^{2}
-E^{2}\left[N^{-1}\left(\hat{t}_{y}-t_{y}\right)\right],
\]
where the expectation is taken over both design, $p_{N}$, and population $%
\xi $ in (\ref{model}).

\begin{theorem}
\label{THM:GJLB} Under Assumptions (A1)-(A8), $\hat{t}_{y,\SBLL}$
asymptotically attains the Godambe-Joshi lower bound, in the sense
that
\[
n_{N}E\left( \frac{\hat{t}_{y,\SBLL}-t_{y}}{N}\right) ^{2}=\frac{n_{N}%
}{N^{2}}\sum_{i\in U_{N}}\sigma^{2}\left( \mathbf{x}_{i}\right) \frac{1-\pi _{i}%
}{\pi _{i}}+o\left( 1\right) .
\]
\end{theorem}

The proofs of Theorems \ref{THM:ADU}-\ref{THM:GJLB} are
given in the Appendix.

\section{Auxiliary Variable Selection}
\label{SEC:varsel}

In this section, we propose a BIC-based method to select the auxiliary
variables for use in the superpopulation model (\ref{model:am}).

The BIC was first proposed in \cite{S78} for the selection of parametric
models. Recently, \cite{HY04} proposed a fast and consistent model selection
method based on spline estimation with the BIC to select significant lags in
non-linear additive autoregression. Analogous to the approach in \cite{HY04},
if the entire realization were known by ``oracle'', one can select significant
auxiliary variables based on the BIC. For an index set of variables $r\in \{1,
..., d\}$, the BIC is defined as
\begin{equation}
\text{BIC}^{\left( r\right) }=\log \left\{ \text{AMSE}^{\left( r\right) }\left(
N^{-1}\hat{t}_{y,\SBLL}\right)\right\} +
\frac{\mathcal{J}_{r}}{n_{N}}\log (n_{N}),
\label{DEF:BIC}
\end{equation}
where $\mathcal{J}_{r}=1+\sum_{\alpha \in r}(J_{N}+1)$, and $\text{AMSE}\left(
N^{-1}\hat{t}_{y,\SBLL}\right)$ is the asymptotic mean squared error (AMSE) of
$N^{-1}\hat{t}_{y,\SBLL}$ in (\ref{DEF:AMSE}), i.e. the asymptotic expectation
of $\left\{{N}^{-1}\left(\hat{t}_{y,\SBLL}-t_{y}\right)\right\} ^{2}$.

Next let $f=n_{N}/N$ be the fixed sampling fraction. Under simple random
sampling (SRS) design, if $\sigma^2(\mathbf{x})=\mathbf{c}^{\T}\mathbf{x}$,
\begin{eqnarray*}
\text{AMSE}\left(
N^{-1}\hat{t}_{y,\SBLL}\right)=\frac{1-f}{n_{N}\left( N-1\right)}
\sum_{i\in U_N}\left( y_{i}-\tilde{m}_{i}^{*}\right) ^{2}.
\end{eqnarray*}
Thus, using similar arguments in Section 5 of \cite{HY04}, we can show that the
above BIC in (\ref{DEF:BIC}) is consistent under SRS.

By Theorem \ref{THM:Vhat-AMSE}, $\text{AMSE}\left(
N^{-1}\hat{t}_{y,\SBLL}\right)$ can be estimated consistently by
\begin{equation}
\widehat{V}_{g}\equiv \widehat{V}_{g}\left( N^{-1}\hat{t}_{y,{\mbox{\rm \tiny SBLL}%
}}\right)=\frac{1}{N^{2}}\sum_{i,j\in
s}\frac{\Delta_{ij}}{\pi _{ij}}\frac{g_{is}\left( y_{i}-\hat{m}
_{i}^{*}\right) }{\pi _{i}}\frac{g_{js}( y_{j}-\hat{m}%
_{j}^{*})}{\pi _{j}}, \label{DEF:Vhatg}
\end{equation}
a modified version of (\ref{DEF:Vhat}) proposed by \cite{SSW89} with the
``g-weight'' in (\ref{DEF:gweight}). So the sample-based BIC is defined as
\begin{equation}
\text{BIC}^{\left( r\right) }=\log \left\{ \widehat{V}_{g}^{\left(
r\right) }\right\} + \frac{\mathcal{J}_{r}}{n_{N}}\log (n_{N}),
\label{DEF:sampleBIC}
\end{equation}
and we select the subsect $ \hat{r} \subset\left\{1,...,d\right\}$ that gives
the smallest BIC value.

\vskip .1in \noindent \textbf{Remark 3.} Under SRS design, the variance
estimator given in (\ref{DEF:Vhatg}) can be simplified as
\[
\widehat{V}_{g}=\frac{1-f}{n_{N}\left( n_{N}-1\right) }\sum_{i\in s}g_{is}^{2}
\left( y_{i}-\hat{m}_{i}^{*}\right) ^{2}.
\]

In practice, we first decide on a set of candidate variables to be selected.
Since a full search through all possible subsets of variables is in general
computationally too costly in actual implementation of the BIC method, we
consider a \textit{forward selection} procedure and a \textit{backward
selection} procedure. Let $d$ denote the total number of candidate variables to
be selected from. In the \textit{forward selection} procedure, we pre-specify
the maximal number of variables $d_{\max}=\min\left\{d,
\left[\frac{n_{N}}{2(J_{N}+1)}\right]\right\}$ that are allowed in the model,
in which $\left[ a\right] $ denotes the integer part of $a$. We start from the
empty set of auxiliary variables, add one variable at a time to the current
model, choosing between the various candidate variables that have not yet been
selected by minimizing BIC in (\ref{DEF:sampleBIC}). The process stops when the
number of variables selected reaches $d_{\max}$. In the \textit{backward
selection} procedure, we start with a set of variables of the maximal size
$d_{\max}$, delete one variable at a time by minimizing the BIC and stop when
no variable remains in the model. If $d_{\max}<d$, we first apply the
\textit{forward selection} procedure, then we start with the maximal set of
variables selected in the last step of the forward stage.

\section{Simulation Study} \label{SEC:simulation}

In this section, simulations are carried out to investigate the
finite-sample performance of $\hat{t}_{y,\SBLL}$. For comparison
we also obtained the results of four other estimators: the HT
estimator which does not make use of the auxiliary population, the
linear regression (LREG) estimator in \citep{SSW92}, the one-step
linear spline (LS) estimator defined by
\begin{equation*}
\hat{t}_{y,\LS}=\sum_{i\in s}(y_{i}-\hat{m}_{i})/\pi _{i}%
+\sum_{i\in U_{N}}\hat{m}_{i},\
\hat{m}_{i}=N^{-1}\hat{t}_{y}+\sum_{\alpha
=1}^{d}\hat{m}_{i\alpha}
\end{equation*}
with $\hat{m}_{i\alpha}\equiv \hat{m}_{\alpha}(x_{i\alpha})$ given in
(\ref{DEF:mhatalpha-xalpha}), and the single-index model-assisted (SIM)
estimator in \citep{W09}. The number of knots $J_{N}$ for the LS and SBLL is
determined by (\ref{DEF:Numberofknots}).

For the superpopulation model (\ref{model}), the following four
additive models (no interactions) were considered:
\[
\begin{array}{ll}
2\text{-dim linear:} & Y=-1+2X_{3}+4X_{6}+\sigma _{0}\varepsilon, \\
2\text{-dim quadratic:} & Y=5.5-6X_{2}+8(X_{2}-.5)^{2}-3X_{10}+32(X_{10}-.5)^{3}+%
\sigma _{0}\varepsilon, \\
3\text{-dim mixed:} & Y=8(X_{2}-.5)^{2}+\exp \left(2
X_{5}-1\right)
+\sin\left\{2\pi (X_{8}-.5)\right\} +\sigma _{0}\varepsilon, \\
5\text{-dim sinusoid:} & Y=2+\sum_{\alpha =1}^{d}\sin
\left\{ 2\pi (X_{\alpha}-.5)\right\} +\frac{\sigma _{0}}{2}
(\sum_{\alpha =1}^{d}X_{\alpha})^{1/2}\varepsilon, \ d=5.
\end{array}
\]

The auxiliary variable vectors $\mathbf{x}_{i}$, $i\in U_{N}$,
were generated from i.i.d. uniform $(0,1)$ random vectors. The
errors $\varepsilon$ were generated from i.i.d. $N\left(
0,1\right) $ with noise level $\sigma_{0}=0.1$, $0.4$. The
population size was $N=1000$. SRS Samples were generated of size
$n_{N}=50,100$ and $200$. For each combination of noise level and
sample size, $1000$ replicated SRS samples were selected from the
same population, the estimators were calculated, and the design
bias and the design mean squared errors were computed empirically.

\setlength{\unitlength}{1cm}
\begin{table}[tb]
\caption{Ratio of MSE of the HT, LREG, LS and SIM estimators to
that of the SBLL estimator and the average computing time of the
SBLL estimator based on $1000$ replications of SRS samples from
four fixed populations of size $N=1000$.}
\label{TAB:ratio}\centering \vskip .05in
\begin{tabular*}{1.0\textwidth}{@{\extracolsep{\fill}}ccrrrrrc}
\hline \multirow{2}{*}{Model} & Error & Sample size &
\multicolumn{4}{c}{MSE Ratio} & SBLL
\\ \cline{4-7}
& $\sigma$ & $n_{N}$ & HT & LREG & LS & SIM & {\textrm{%
(seconds)}} \\ \hline\hline
\multirow{6}{*}{$1$} & \multirow{3}{*}{$0.1$} %
   & $50$ & $140.36$ & $0.89$ & $1.12$ & $1.60$ & $0.07$ \\
&  & $100$ & $148.03$ & $0.91$ & $1.07$ & $1.33$ & $0.07$ \\
&  & $200$ & $147.03$ & $0.92$ & $1.10$ & $1.02$ & $0.09$ \\
\cline{2-8} & \multirow{3}{*}{$0.4$} %
   & $50$ & $9.78$ & $0.92$ & $1.16$ & $1.24$ & $0.07$ \\
&  & $100$ & $10.50$ & $0.95$ & $1.10$ & $1.02$ & $0.07$ \\
&  & $200$ & $10.47$ & $0.98$ & $1.05$ & $1.04$ & $0.09$ \\
\hline
\multirow{6}{*}{$2$} & \multirow{3}{*}{$0.1$} %
   & $50$ & $134.05$ & $28.38$ & $2.11$ & $19.77$ & $0.07$ \\
&  & $100$ & $282.47$ & $58.10$ & $1.03$ & $36.58$ & $0.07$ \\
&  & $200$ & $313.93$ & $66.63$ & $0.98$ & $41.15$ & $0.09$ \\
\cline{2-8} & \multirow{3}{*}{$0.4$} %
   & $50$ & $18.45$ & $4.25$ & $2.36$ & $3.44$ & $0.07$\\
&  & $100$ & $23.67$ & $5.34$ & $1.04$ & $3.69$ & $0.07$ \\
&  & $200$ & $23.36$ & $5.63$ & $1.02$ & $3.92$ & $0.09$ \\
\hline
\multirow{6}{*}{$3$} & \multirow{3}{*}{$0.1$} %
& $50$ & $63.14$ & $30.83$ & $1.10$ & $37.12$ & $0.07$ \\
&  & $100$ & $103.33$ & $49.62$ & $1.01$ & $50.76$ & $0.07$ \\
&  & $200$ & $115.13$ & $56.57$ & $1.02$ & $57.04$ & $0.09$ \\
\cline{2-8} & \multirow{3}{*}{$0.4$} %
   & $50$ & $6.80$ & $3.46$ & $1.11$ & $3.93$ & $0.07$ \\
&  & $100$ & $8.18$ & $4.20$ & $1.14$ & $4.40$ & $0.07$ \\
&  & $200$ & $18.39$ & $4.52$ & $1.09$ & $4.57$ & $0.09$ \\
\hline
\multirow{6}{*}{$4$} & \multirow{3}{*}{$0.1$} %
   & $50$ & $55.81$ & $25.26$ & $1.01$ & $27.61$ & $0.07$ \\
&  & $100$ & $151.59$ & $62.63$ & $1.03$ & $65.78$ & $0.07$ \\
&  & $200$ & $230.44$ & $97.91$ & $0.97$ & $99.45$ & $0.09$ \\
\cline{2-8} & \multirow{3}{*}{$0.4$} %
   & $50$ & $9.97$ & $4.75$ & $1.03$ & $5.22$ & $0.07$\\
&  & $100$ & $16.35$ & $7.10$ & $1.01$ & $7.44$ & $0.07$ \\
&  & $200$ & $19.95$ & $8.60$ & $1.05$ & $8.74$ & $0.09$ \\
\hline
\end{tabular*}
\end{table}

\setlength{\unitlength}{1cm}
\begin{table}[tb]
\caption{Monte Carlo bias, standard error and
the square root of the average estimated variances
(\ref{DEF:Vhat}) of the population total based on 1000
simulations.} \label{TAB:variance}\centering \vskip .05in
\begin{tabular*}{1.0\textwidth}{@{\extracolsep{\fill}}ccrrrr} \hline Model &
$\sigma$ & $n$ & Bias & SE & Est. SE \\ \hline
\multirow{6}{*}{$1$} & \multirow{3}{*}{$0.1$} %
   & $50$ & $-0.10$ & $14.69$ & $13.18$\\
&  & $100$ & $-0.36$ & $9.85$ & $9.32$\\
&  & $200$ & $-0.13$ & $6.55$ & $6.29$\\
& \multirow{3}{*}{$0.4$}
   & $50$ & $-1.62$ & $57.73$ & $51.81$\\
&  & $100$ & $-1.55$ & $38.51$ & $36.77$\\
&  & $200$ & $-0.42$ & $25.71$ & $24.86$\\ \hline
\multirow{6}{*}{$2$} & \multirow{3}{*}{$0.1$}%
   & $50$ & $1.27$ & $24.49$ & $14.06$ \\%
&  & $100$ & $0.62$ & $11.52$ & $9.13$\\
&  & $200$ & $0.37$ & $7.06$ & $6.10$\\
& \multirow{3}{*}{$0.4$} & $50$ & $2.41$ & $67.66$ & $52.45$\\
&  & $100$ & $-0.47$ & $40.94$ & $36.15$\\
&  & $200$ & $-0.13$ & $26.54$ & $24.33$\\ \hline
\multirow{6}{*}{$3$} & \multirow{3}{*}{$0.1$} %
   & $50$ & $2.29$ & $20.40$ & $13.38$\\
&  & $100$ & $0.90$ & $10.91$ & $8.74$\\
&  & $200$ & $0.48$ & $6.82$ & $5.88$\\
& \multirow{3}{*}{$0.4$}
   & $50$ & $2.17$ & $64.89$ & $50.86$\\
&  & $100$ & $-0.04$ & $40.17$ & $35.44$\\
&  & $200$ & $0.32$ & $26.30$ & $23.99$\\ \hline
\multirow{6}{*}{$4$} & \multirow{3}{*}{$0.1$}%
   & $50$ & $-1.98$ & $29.04$ & $18.04$ \\%
&  & $100$ & $-0.51$ & $12.28$ & $8.22$\\
&  & $200$ & $-0.10$ & $6.38$ & $4.82$\\
& \multirow{3}{*}{$0.4$} & $50$ & $-4.38$ & $69.69$ & $43.31$\\
&  & $100$ & $-1.18$ & $37.92$ & $27.72$\\
&  & $200$ & $-0.37$ & $22.58$ & $18.56$\\ \hline
\end{tabular*}
\end{table}

Table \ref{TAB:ratio} shows the ratios of the mean squared
error (MSE) for the various estimators to the proposed SBLL
estimators. From the table, one sees that the model-assisted
estimators, LREG, LS, SIM and SBLL, perform much better
than the simple HT regardless the type of mean
function, standard error and sample size. For Model $1$, LREG is
expected to be the preferred estimator, since the assumed model is
correctly specified. However, not much efficiency is lost by using
SBLL instead of LREG and the MSE ratios of LREG to SBLL are at
least $0.89$ for all cases. For all other scenarios, SBLL performs
consistently better than LREG. The SBLL estimators also improve
upon the LS estimators across almost every combination of noise
level and sample size, which implies that our second local linear
smoothing step is not redundant.

\setlength{\unitlength}{1cm}
\begin{table}[tb]
\caption{Simulation results for auxiliary variable selection based
on $100$ replications of SRS samples from four fixed populations
of size $N=1000$. (Here the MSE Ratio is the ratio of MSE of the
SBLL estimator calculated by using the selected model to the MSE
of the oracle SBLL estimates computed by using the true model.)}
\label{TAB:selection}\centering \vskip .05in
\begin{tabular*}{1.0\textwidth}{@{\extracolsep{\fill}}ccr|rrrcrrrc}
\hline \multirow{2}{*}{Model} & \multirow{2}{*}{$\sigma_0$} &
\multirow{2}{*}{$n$} & \multicolumn{4}{c}{Forward} & \multicolumn{4}{c}{Backward}\\
\cline{4-7} \cline{8-11} & & & \multirow{2}{*}{C} &
\multirow{2}{*}{U} & \multirow{2}{*}{O} & MSE &
\multirow{2}{*}{C} & \multirow{2}{*}{U} & \multirow{2}{*}{O}
& MSE\\
& & & & & & Ratio & & & & Ratio\\
\hline \multirow{6}{*}{$1$} & \multirow{3}{*}{$0.1$}%
  & $50$  & $72$ & $0$ & $28$ & $1.150$ & $73$& $0$ & $27$ & $1.124$ \\
& & $100$ & $97$ & $0$ & $3$ & $1.001$ & $97$ & $0$ & $3$ & $1.001$\\
& & $200$ & $99$ & $0$ & $1$ & $0.999$ & $99$ & $0$ & $1$ & $0.999$\\
\cline{2-11} & \multirow{3}{*}{$0.4$} %
  & $50$ & $76$ & $0$ & $24$ & $1.147$ & $77$ & $0$ & $23$ & $1.145$\\
& & $100$ & $98$ & $0$ & $2$ & $1.002$ & $98$ & $0$ & $2$ & $1.002$\\
& & $200$ & $100$ & $0$ & $0$ & $1.000$ & $100$ & $0$ & $0$ & $1.000$\\
\hline \multirow{6}{*}{$2$} & \multirow{3}{*}{$0.1$} %
  & $50$ & $87$ & $0$ & $13$ & $1.255$ & $87$ & $0$ & $13$ & $1.255$\\
& & $100$ & $96$ & $0$ & $4$ & $1.012$ & $96$ & $0$ & $4$ & $1.012$\\
& & $200$ & $100$ & $0$ & $0$ & $1.000$ & $100$ & $0$ & $0$ & $1.000$\\
\cline{2-11} & \multirow{3}{*}{$0.4$} %
  & $50$ & $79$ & $0$ & $21$ & $1.019$ & $80$ & $0$ & $20$ & $1.022$\\
& & $100$ & $98$ & $0$ & $2$ &$1.000$ & $98$ & $0$ & $2$ & $1.000$\\
& & $200$ & $100$ & $0$ & $0$ & $1.000$ & $100$ & $0$ & $0$ & $1.000$\\
\hline \multirow{6}{*}{$3$} & \multirow{3}{*}{$0.1$}
  & $50$ & $87$ & $0$ & $13$ & $1.082$ & $86$ & $0$ & $14$ & $1.082$\\
& & $100$ & $91$ & $0$ & $9$ & $1.000$ & $91$ & $0$ & $9$ & $1.001$\\
& & $200$ & $100$ & $0$ & $0$ & $1.000$ & $100$ & $0$ & $0$ & $1.000$\\
\cline{2-11} & \multirow{3}{*}{$0.4$} %
  & $50$ & $83$ & $0$ & $17$ & $1.020$ & $83$ & $0$ & $17$ & $1.020$\\
& & $100$ & $99$ & $0$ & $1$ & $1.000$ & $99$ & $0$ & $1$ & $1.000$\\
& & $200$ & $100$ & $0$ & $0$ & $1.000$ & $100$ & $0$ & $0$ & $1.000$\\
\hline \multirow{6}{*}{$4$} & \multirow{3}{*}{$0.1$} %
  & $50$ & $68$ & $0$ & $32$ & $1.277$ & $69$ & $0$ & $31$ & $1.277$\\
& & $100$ & $88$ & $0$ & $12$ & $1.029$ & $88$ & $0$ & $12$ & $1.029$\\
& & $200$ & $100$ & $0$ & $0$ & $1.000$ & $100$ & $0$ & $0$ & $1.000$\\
\cline{2-11} & \multirow{3}{*}{$0.4$} %
  & $50$ & $69$ & $0$ & $31$ & $1.063$ & $69$ & $0$ & $31$ & $1.063$\\
& & $100$ & $97$ & $0$ & $3$ & $1.000$ & $97$ & $0$ & $3$ & $1.031$\\
& & $200$ & $100$ & $0$ & $0$ & $1.000$ & $100$ & $0$ & $0$ & $1.000$\\
\hline
\end{tabular*}
\end{table}

To see how fast the computation is, Table \ref{TAB:ratio} also
provides the average time of generating one sample of size $n_{N}$
and obtaining the SBLL estimator on an ordinary PC with Intel
Pentium IV 1.86 GHz processor and 1.0 GB RAM. It shows that the
proposed SBLL estimation is extremely fast. For instance, for
Model $4$, the SBLL estimation of a $5$-dimensional of size $200$
takes on average merely $0.2$ second. We also carried out
simulations for high dimensional data with sample size
$n_{N}=1000$ generated from the population of size $10000$.
Remarkably, it takes on average less than $60$ seconds to get the
SBLL estimator even when the dimension reaches $50$.

In Table \ref{TAB:variance} we give the Monte
Carlo bias and standard error of the SBLL estimator based on
its sampling distribution over $1000$ replications. Table
\ref{TAB:variance} also show the square root of the average
estimated variance of the population total (\ref{DEF:Vhat}). We
see that the biases of the SBLL estimator are very small and the
variance estimator appears to perform well for medium sample
size.

Next we conducted simulations to evaluate the performance of the variable
selection method. We generated $100$ replications for each of the above models.
The variables were searched from $\left\{1,2,...,10\right\}$ for all methods
and we set the maximum number of variables allowed in the model to be $10$.
Table \ref{TAB:selection} shows the number of correct fit (C), underfit (U) and
overfit (O) based on the BIC in (\ref{DEF:sampleBIC}) over 100 simulation runs.
Here underfitting means that the method misses at least one of the significant
variables. From Table \ref{TAB:selection}, we can see that both the forward and
the backward selection procedures perform very well for moderately large sample
size. We also obtained the ratio of MSE of the SBLL estimates calculated by
using the selected model to the MSE of the oracle SBLL estimates computed by
using the true model. In all the cases, the ratios are very close to 1 or
exactly 1 for moderately large sample size.

\section{Discussion} \label{SEC:discussion}

Nonparametric additive methods enhance the flexibility of the models that
survey practitioners use. However, due to the limitations in either
interpretability, computational complexity or theoretical reliability, these
models have not been widely used as general tools in survey data analysis. In
this paper, we have advanced additive models as flexible, computationally
efficient and theoretically attractive tools for studying survey data. We also
developed a consistent procedure to select the significant auxiliary variables
under simple random sampling design.

The proposed method in this paper is appropriate only for survey data that
follow simple additive model. The limitation of the basic additive model is
that the interactions between the input features are not considered. There are
other models, for instance, single-index model \citep{W09}, additive model with
second-order interaction terms \citep{STY02}, which reduce dimensionality but
also incorporate interactions. Additive partially linear model \citep{FL03} is
another parsimonious candidate when one believes that the relationship between
the study variable and some of the auxiliary variables has a parametric form,
while the relationship between the study variable and the remaining auxiliary
covariates may not be linear. These alternative models are supposed to be more
efficient in certain cases, but obtaining the asymptotics is likely to be very
complicated, thus we leave it as future research work.

Finally, in our methodology development, we have assumed that the auxiliary
variables are available for all population elements. It would be interesting to
consider the limited auxiliary information case \citep{CDW98} where only some
summary quantities such as means are available at the population level. This is
also a challenging problem for future research.

\vspace{0.5cm} 
\noindent \textbf{Appendix} \vspace{.2cm}

\renewcommand{\thetheorem}{{\sc
A.\arabic{theorem}}}
\renewcommand{\theproposition}{{\sc
A.\arabic{proposition}}} \renewcommand{\thelemma}{{\sc A.\arabic{lemma}}} %
\renewcommand{\thecorollary}{{\sc A.\arabic{corollary}}} \renewcommand{%
\theequation}{A.\arabic{equation}} \renewcommand{\thesubsection}{A.%
\arabic{subsection}} 
\setcounter{theorem}{0}

To show the asymptotic properties of the proposed estimator
$\hat{t}_{y,\SBLL}$, we first introduce an ``oracle'' SBLL
estimator of $t_{y}$ if the entire realization were known.

\vspace{.2cm} \setcounter{equation}{0} 
\noindent \textbf{A.1. The Population-based
Estimator}

If the entire realization were known, let $\mathbf{\Gamma} _{U}=\left\{
\mathbf{\Gamma }\left(\mathbf{x}_{i} \right)^{\T}\right\}_{i\in U_{N}}$ be the
population-based truncated power spline matrix, where $\mathbf{\Gamma
}\left(\mathbf{x} \right)$ is given in (\ref{DEF:Gamma(x)}). Let $\mathbf{y}$
be the vector of the response values $y_{i}$ for $i\in U_{N}$. Further let
$\mathbf{B}_{U}=\left( \mathbf{\Gamma }_{U}^{\T}\mathbf{\Gamma }_{U}\right)
^{-1}\mathbf{\Gamma }_{U}^{\T}\mathbf{y}$. The centered pilot estimators of
$m_{\alpha }\left( x_{\alpha }\right)$ at the first stage is
\begin{equation}
\tilde{m}_{\alpha }\left( x_{\alpha }\right) =\mathbf{\Gamma
}\left(\mathbf{x}\right)^{\T}\mathbf{D}_{\alpha}\mathbf{B}_{U}-N^{-1}\mathbf{1}_{N}^{\T}%
\mathbf{\Gamma }_{U}\mathbf{D}_{\alpha }\mathbf{B}_{U},
\label{DEF:mtildealpha-xalpha}
\end{equation}
where vector $\mathbf{1}_{N}^{\T}=\left\{1,1,...,1\right\}$ of
length $N$. The pilot estimators for all elements in the
population is denoted by
\[
\tilde{\mathbf{m}}_{\alpha }\equiv \left\{ \tilde{m}_{\alpha
}\left(
x_{i\alpha }\right) \right\} _{i\in U_{N}}=\left( \mathbf{I}-N^{-1}\mathbf{1}%
_{N}\mathbf{1}_{N}^{\T}\right) \mathbf{\Gamma }_{U}\mathbf{D}_{\alpha }\mathbf{B}%
_{U},\mbox{\quad} \alpha =1,...,d.
\]
For the second stage kernel smoothing, define the matrices
\[
\mathbf{X}_{Ui\alpha }=\left\{\left(
\begin{array}{ll}
1 & x_{k\alpha }-x_{i\alpha }
\end{array}
\right) \right\}_{k\in U_{N}},\mbox{\quad} \mathbf{W}_{Ui\alpha
}=\text{diag}\left\{ K_{h}\left( x_{k\alpha }-x_{i\alpha }\right)
\right\} _{k\in U_{N}}.
\]
Then the SBLL estimator of each component at $\mathbf{x}_{i}$ is
given by
\begin{equation}
\tilde{m}_{i\alpha }^{*}\equiv \mathbf{e}_{1}^{\T}\left(
\mathbf{X}_{Ui\alpha
}^{\T}\mathbf{W}_{Ui\alpha }\mathbf{X}_{Ui\alpha }\right) ^{-1}\mathbf{X}%
_{Ui\alpha }^{\T}\mathbf{W}_{Ui\alpha }\tilde{\mathbf{y}}_{\alpha
}, \label{DEF:mtildestar-ialpha}
\end{equation}
where
$\tilde{\mathbf{y}}_{\alpha}=\mathbf{y}-\frac{1}{N}t_{y}\mathbf{1}_{N}-\sum_{\beta
\neq \alpha }\tilde{\mathbf{m}}_{\beta }$ is collection of the
pseudo-responses. The SBLL estimator of $m\left(\mathbf{x}_{i}\right)$ based on
the entire population is given by
\begin{equation}
\tilde{m}_{i}^{*}=\frac{1}{N}t_{y}+\sum_{\alpha
=1}^{d}\tilde{m}_{i\alpha }^{*}, \mbox{\quad} i\in U_{N}.
\label{DEF:mitildestar}
\end{equation}
Clearly, $\tilde{m}_{i}^{*}$ is the prediction at $\mathbf{x}_{i}$
based on the entire finite population. If these $\tilde{m}_{i}^{*}$ were known, a design-unbiased estimator of
$t_{y}$ would be
\begin{equation}
\tilde{t}_{y,\SBLL}=\sum_{i\in
U_{N}}\tilde{m}_{i}^{*}+\sum_{i\in
s}\frac{y_{i}-\tilde{m}_{i}^{*}}{\pi _{i}}.  \label{DEF:tySBLLtilde}
\end{equation}

The proof of the asymptotic properties of $\hat{t}_{y,\SBLL}$ uses
reasoning similar to that in \citep{BO00}, in which a
key step is the Taylor linearization. Recall that our proposed
estimator involves two smoothing stages: spline smoothing in the
first stage and kernel smoothing in the second stage. In the
following, we establish the Taylor linearization for these two
smoothing stages one by one.

\vspace{.2cm}
\noindent \textbf{A.2. Taylor Linearization at the
First Stage}

\begin{lemma}
\label{LEM:sup(mtilde-hat)} Under Assumptions (A1)-(A7), for any
$\alpha =1,...,d$,
\[
\lim_{N\rightarrow \infty }\sup_{x_{\alpha }\in \left[ 0,1\right]
}\left| \hat{m}_{\alpha }\left( x_{\alpha }\right)
-\tilde{m}_{\alpha }\left( x_{\alpha }\right) \right| =O_{p}\left\{
J_{N}(N^{-1}{\log N})^{1/2}\right\} ,
\]
where $\hat{m}_{\alpha }\left( x_{\alpha }\right) $ and $\tilde{m}_{\alpha
}\left( x_{\alpha }\right) $ are the pilot estimators given in
(\ref{EQ:mhatalpha-xalpha}) and (\ref{DEF:mtildealpha-xalpha}).
\end{lemma}

\noindent \textbf{Proof.} Let $\mathbf{S}=N^{-1}\mathbf{\Gamma }_{U}^{{%
\mbox{\rm\tiny T}}}\mathbf{\Gamma }_{U}$ and $\mathbf{V}=N^{-1}\mathbf{%
\Gamma }_{U}^{{\mbox{\rm\tiny T}}}\mathbf{y}$ be matrices with
components $s_{jj^{\prime }}=N^{-1}\sum_{k\in U_{N}}\Gamma_{U,kj}
\Gamma_{U,kj^{\prime }}$ and $v_{j}=N^{-1}\sum_{k\in
U_{N}}\Gamma_{U,kj}y_{k} $, respectively. Denote $\mathbf{S}_{\pi }=N^{-1}%
\mathbf{\Gamma }_{s}^{{\mbox{\rm\tiny T}}}\mathbf{\Pi }_{s}\mathbf{\Gamma }%
_{s}$ and $\mathbf{V}_{\pi }=N^{-1}\mathbf{\Gamma }_{s}^{{\mbox{\rm\tiny T}}}%
\mathbf{\Pi }_{s}\mathbf{y}_{s}$ the sample versions of the matrices $%
\mathbf{S}$ and $\mathbf{V}$ with components $s_{\pi ,jj^{\prime
}}=N^{-1}\sum_{k\in s}\left. \Gamma_{s,kj}\Gamma_{s,kj^{\prime
}}\right/ \pi _{k}$ and $v_{\pi ,j}=N^{-1}\sum_{k\in s}\left.
\Gamma_{s,kj}y_{k}\right/
\pi _{k}$. For each $\alpha =1,...,d$ and the spline basis $\mathbf{\Gamma}%
\left(\mathbf{x}\right) $ in (\ref{DEF:Gamma(x)}), let
\begin{equation}
\zeta \left( \mathbf{S}_{\pi },\mathbf{V}_{\pi };x_{\alpha }\right) =\mathbf{%
\Gamma } \left( \mathbf{x}\right)^{{\mbox{\rm\tiny T}}}
\mathbf{D}_{\alpha
}\left( \mathbf{S}_{\pi }^{-1}\mathbf{V}_{\pi }-\mathbf{S}^{-1}\mathbf{V}%
\right)  \label{DEF:zeta}
\end{equation}
be a nonlinear function of $\left\{ s_{\pi ,jj^{\prime }}\right\} _{1\leq
j,j^{\prime }\leq G_{d}}$ and $\left\{ v_{\pi ,j}\right\} _{j=1}^{G_{d}}$ with
respect to $x_{\alpha }$. The difference $\hat{m}_{\alpha }\left( x_{\alpha
}\right) -\tilde{m}_{\alpha }\left( x_{\alpha }\right) =\zeta \left(
\mathbf{S}_{\pi },\mathbf{V}_{\pi };x_{\alpha }\right)+O_{p}(N^{-1/2})$. Simple
calculation shows that the first order derivatives of $\zeta $ in
(\ref{DEF:zeta}) of $s_{\pi ,jj^{\prime }}$ and $v_{\pi ,j}$ are
\begin{eqnarray*}
\frac{\partial \zeta }{\partial s_{\pi ,jj^{\prime
}}}&=&\mathbf{\Gamma}
\left( \mathbf{x}\right) ^{{\mbox{\rm\tiny T}}}\mathbf{D}_{\alpha }\left( -%
\mathbf{S}_{\pi }^{-1}\mathbf{\Lambda }_{jj^{\prime
}}\mathbf{S}_{\pi }^{-1}\right) \mathbf{V}_{\pi },\quad 1\leq
j,j^{\prime }\leq G_{d},\\
\frac{\partial \zeta }{\partial v_{\pi ,j}}&=&\mathbf{\Gamma} \left( \mathbf{x}%
\right) ^{{\mbox{\rm\tiny T}}}\mathbf{D}_{\alpha }\mathbf{S}_{\pi }^{-1}{%
\mbox{\boldmath $\lambda$}}_{j},\quad 1\leq j\leq G_{d},
\end{eqnarray*}
where ${\mbox{\boldmath $\lambda$}}_{j}$ is a $G_{d}$-vector with
``1'' in the $j$th component and ``0'' elsewhere; and
$\mathbf{\Lambda }_{jj^{\prime }}$ is a $G_{d} \times G_{d}$
matrix with ``1'' in positions $\left( j,j^{\prime }\right) $ and
$\left( j^{\prime },j\right) $ and ``0'' everywhere else.

Using the Taylor linearization, one can approximate $\zeta $ in
(\ref {DEF:zeta}) by a linear one so that the difference between
$\hat{m}_{\alpha }\left( x_{\alpha }\right) $ and
$\tilde{m}_{\alpha }\left( x_{\alpha }\right) $ can be decomposed as
$
\sum_{j=1}^{G_{d}}\varphi _{\alpha j}\left( x_{\alpha }\right)
\left( v_{\pi ,j}-v_{j}\right) -\hat{c}_{\alpha
}+\tilde{c}_{\alpha} -\sum_{1\leq j,j^{\prime }\leq G_{d}}\psi
_{\alpha jj^{\prime }}\left( x_{\alpha }\right) \left( s_{\pi
,jj^{\prime }}-s_{jj^{\prime }}\right) +Q_{\alpha N}\left(
x_{\alpha }\right)
$,
where for any $1\leq j,j^{\prime }\leq G_{d}$,
\begin{eqnarray*}
\varphi _{\alpha j}\left( x_{\alpha }\right) &=&\left.
\frac{\partial \zeta }{\partial v_{\pi ,j}}\right| _{v_{\pi
,j}=v_{j}}=\mathbf{\Gamma} \left(
\mathbf{x}\right) ^{{\mbox{\rm\tiny T}}}\mathbf{D}_{\alpha }\mathbf{S}^{-1}%
\mathbf{\lambda }_{j}, \\
\psi _{\alpha jj^{\prime }}\left( x_{\alpha }\right) &=&\left. \frac{%
\partial \zeta }{\partial s_{\pi ,jj^{\prime }}}\right| _{s_{\pi ,jj^{\prime
}}=s_{jj^{\prime }}}=\mathbf{\Gamma} \left( \mathbf{x}\right) ^{{%
\mbox{\rm\tiny T}}}\mathbf{D}_{\alpha }\left(
\mathbf{S}^{-1}\mathbf{\Lambda }_{jj^{\prime
}}\mathbf{S}^{-1}\right) \mathbf{V},
\end{eqnarray*}
and $Q_{\alpha N}\left( x_{\alpha }\right) $ is the remainder.
Note that
\begin{eqnarray*}
&&\sum_{j=1}^{G_{d}}\varphi _{\alpha j}\left( x_{\alpha }\right)
\left( v_{\pi ,j}-v_{j}\right) = N^{-1}\sum_{k\in
U_{N}}\sum_{j=1}^{G_{d}}\varphi
_{\alpha j}\left( x_{\alpha }\right) \Gamma_{U,kj}y_{k}\left( 1-\frac{I_{k}}{%
\pi _{k}}\right) \\
&&-N^{-1}\sum_{k\in U_{N}}\sum_{j=1}^{G_{d}}\varphi _{\alpha
j}\left(
x_{\alpha }\right) \left(\Gamma_{U,kj}-\Gamma_{s,kj}\right) \left( 1-\frac{%
I_{k}}{\pi _{k}}\right) y_{k}\\
&&+N^{-1}\sum_{k\in
U_{N}}\sum_{j=1}^{G_{d}}\varphi _{\alpha j}\left( x_{\alpha
}\right) y_{k}\left(\Gamma_{U,kj}-\Gamma_{s,kj}\right) .
\end{eqnarray*}
By the discretization method given in Lemma A.4 of \citep{WY07},
the Borel-Cantelli Lemma entails that each single term in the
right hand side of the above is of the order
$O_{p}\left\{J_{N}(N^{-1}{\log N})^{1/2}\right\}$. Therefore, we
have
\[
\sup_{x_{\alpha }\in \left[ 0,1\right] }\left|
\sum_{j=1}^{G_{d}}\varphi _{\alpha j}\left( x_{\alpha }\right)
\left( v_{\pi ,j}-v_{j}\right) \right|
=O_{p}\left\{
J_{N}(N^{-1}{\log N})^{1/2}\right\}.
\]
Similar
arguments lead to $\sup_{x_{\alpha }\in \left[ 0,1\right] }\left| \sum_{1\leq
j,j^{\prime }\leq G_{d}}\psi _{\alpha jj^{\prime }}\left(
x_{\alpha }\right) \left( s_{\pi
,jj^{\prime }}-s_{jj^{\prime }}\right) \right|$ is of the order
$O_{p}\left\{J_{N}(N^{-1}{\log N})^{1/2}\right\}$,
and $\sup_{x_{\alpha }\in \left[ 0,1\right] }\left| Q_{\alpha
N}\left( x_{\alpha }\right) \right|
=o_{p}\left\{
J_{N}(N^{-1}{\log N})^{1/2}\right\}$.
Thus $\sup_{x_{\alpha }\in \left[ 0,1\right] }\zeta \left( \mathbf{S}_{\pi },%
\mathbf{V}_{\pi };x_{\alpha }\right) =O_{p}\left\{
J_{N}(N^{-1}{\log N})^{1/2}\right\}$. The desired result is established. \hfill $%
\square $

\vspace{.2cm}
\noindent \textbf{A.3. Taylor Linearization at the Second Stage} \vspace{.2cm}

Let
\begin{eqnarray*}
t_{i\alpha q} &=&\sum_{k\in U_{N}}K_{h}\left( x_{k\alpha
}-x_{i\alpha
}\right) \left( x_{k\alpha }-x_{i\alpha }\right) ^{q-1},\\ \hat{t}%
_{i\alpha q}&=&\sum_{k\in s}\frac{1}{\pi
_{k}}K_{h}\left( x_{k\alpha }-x_{i\alpha
}\right) \left( x_{k\alpha }-x_{i\alpha }\right) ^{q-1},
\end{eqnarray*}
for $q=1,2,3$ and
\begin{eqnarray*}
t_{i\alpha q} &=& \sum_{k\in U_{N}}K_{h}\left( x_{k\alpha
}-x_{i\alpha }\right) \left( x_{k\alpha }-x_{i\alpha
}\right)^{q-4} \tilde{y}_{k\alpha }, \\ \hat{t}_{i\alpha
q}&=&\sum_{k\in s}\frac{1}{\pi _{k}}K_{h}\left( x_{k\alpha }-x_{i\alpha
}\right) \left( x_{k\alpha }-x_{i\alpha }\right)^{q-4} \hat{y}_{k\alpha },
\end{eqnarray*}
for $q=4,5$. We rewrite $\tilde{m}_{i\alpha }^{*}$ in (\ref
{DEF:mtildestar-ialpha}) and $\hat{m}_{i\alpha }^{*}$ in (\ref
{DEF:mhatstar-ialpha}) by
\[
\tilde{m}_{i\alpha }^{*}=\frac{t_{i\alpha 3}t_{i\alpha
4}-t_{i\alpha 2}t_{i\alpha 5}}{t_{i\alpha 1}t_{i\alpha
3}-t_{i\alpha 2}^{2}}, \mbox{\quad}
\hat{m}_{i\alpha }^{*}=\frac{\hat{t}_{i\alpha 3}\hat{t}_{i\alpha 4}-\hat{t}%
_{i\alpha 2}\hat{t}_{i\alpha 5}}{\hat{t}_{i\alpha 1}\hat{t}_{i\alpha 3}-\hat{%
t}_{i\alpha 2}^{2}}.
\]
Let $z_{i\alpha k}=\sum_{q=1}^{5}\left. \frac{\partial \hat{m}_{i\alpha }^{*}%
}{\partial \left( N^{-1}\hat{t}_{i\alpha q}\right) }\right| _{\hat{\mathbf{t}%
}_{i\alpha}=\mathbf{t}_{i\alpha}}z_{i\alpha kq}$, where $\mathbf{t}%
_{i\alpha}=\left\{t_{i\alpha q}\right\}_{q=1}^{5}$ and
\[
z_{i\alpha kq} = \left\{
\begin{array}{ll}
K_{h}\left( x_{k\alpha }-x_{i\alpha }\right) \left( x_{k\alpha
}-x_{i\alpha
}\right) ^{q-1}, & \text{for } q=1,2,3, \\
K_{h}\left( x_{k\alpha }-x_{i\alpha }\right) \left( x_{k\alpha
}-x_{i\alpha }\right)^{q-4} y_{k\alpha }, & \text{for } q=4,5.
\end{array}
\right.
\]
Then one can approximate $\hat{m}_{i\alpha
}^{*}-\tilde{m}_{i\alpha }^{*}$ by a linear sum, i.e.,
\begin{equation}
\hat{m}_{i\alpha }^{*}-\tilde{m}_{i\alpha
}^{*}=\frac{1}{N}\sum_{k\in U_{N}}z_{i\alpha k}\left(
\frac{I_{k}}{\pi _{k}}-1\right) -L_{i\alpha N}+R_{i\alpha N},
\label{EQ:Taylor expansion2}
\end{equation}
where $L_{i\alpha N}=\sum_{q=1}^{4}L_{i\alpha Nq}$ with
\begin{eqnarray*}
&& \hspace{-0.7cm} L_{i\alpha N1} = \frac{1}{N^{2}}\left( \hat{t}%
_{y}-t_{y}\right) \left. \frac{\partial \hat{m}_{i\alpha
}^{*}}{\partial
\left( N^{-1}\hat{t}_{i\alpha 4}\right) }\right|_{\hat{\mathbf{t}}_{i\alpha}=%
\mathbf{t}_{i\alpha}}\sum_{k\in U_{N}}z_{i\alpha k1}\left(
\frac{I_{k}}{\pi
_{k}}-1\right) , \\
&& \hspace{-0.7cm} L_{i\alpha N2} = \frac{1}{N}\left. \frac{\partial \hat{m}%
_{i\alpha }^{*}}{\partial \left( N^{-1}\hat{t}_{i\alpha 4}\right) }\right|_{%
\hat{\mathbf{t}}_{i\alpha}=\mathbf{t}_{i\alpha}}\sum_{k\in
U_{N}}z_{i\alpha k1}\left( \frac{I_{k}}{\pi _{k}}-1\right)
\sum_{\beta \neq \alpha }\left\{ \hat{m}_{\beta }\left( x_{k\beta
}\right) -\tilde{m}_{\beta }\left( x_{k\beta }\right) \right\},\\
&& \hspace{-0.7cm} L_{i\alpha N3} = \frac{1}{N^{2}}\left( \hat{t}%
_{y}-t_{y}\right) \left. \frac{\partial \hat{m}_{i\alpha
}^{*}}{\partial
\left( N^{-1}\hat{t}_{i\alpha 5}\right) }\right| _{\hat{\mathbf{t}}%
_{i\alpha}=\mathbf{t}_{i\alpha}}\sum_{k\in U_{N}}z_{i\alpha k2}\left( \frac{%
I_{k}}{\pi _{k}}-1\right) , \\
&& \hspace{-0.7cm} L_{i\alpha N4} =\frac{1}{N}\left. \frac{\partial \hat{m}%
_{i\alpha }^{*}}{\partial \left( N^{-1}\hat{t}_{i\alpha 5}\right) }\right|_{%
\hat{\mathbf{t}}_{i\alpha}=\mathbf{t}_{i\alpha}}\sum_{k\in
U_{N}}z_{i\alpha k2}\left( \frac{I_{k}}{\pi _{k}}-1\right)
\sum_{\beta \neq \alpha } \left\{ \hat{m}_{\beta }\left( x_{k\beta
}\right) -\tilde{m}_{\beta }\left( x_{k\beta }\right) \right\} ,
\end{eqnarray*}
and $R_{i\alpha N}$ is the remainder. Similar to the proof of
Lemma 3 in \citep{BO00},
\begin{equation}
\frac{n_{N}}{N}\sum_{i\in U_{N}}E_{p}\left[ R_{i\alpha
N}^{2}\right] =O\left(n_{N}^{-1}h_{N}^{-2}\right) .
\label{EQ:remainder}
\end{equation}

\begin{lemma}
\label{LEM:Remainders} Under Assumptions (A1)-(A8), $N^{-1}\sum_{i\in U_{N}}E_{p}\left(
L_{i\alpha N}^{2}\right) \rightarrow 0$.
\end{lemma}

\noindent \textbf{Proof.} By the Cauchy-Schwartz inequality, it
suffices to show that for $q=1,...,4$, $N^{-1}\sum_{i\in U_{N}}E_{p} \left(
L_{i\alpha Nq}^{2}\right) \rightarrow 0$.
Without loss of generality, we only show the cases for $q=1$ and
$2$. Similarly to the proof of Lemma 2 (v) in \citep{BO00}, the first order derivatives of $\hat{m}_{i\alpha
}^{*}$ with respect to $N^{-1}\hat{t}_{i\alpha q}$ evaluated at $\hat{%
\mathbf{t}}_{i}=\mathbf{t}_{i}$ are uniformly bounded in $i$. So
by Assumption (A7)
\begin{eqnarray*}
&&  \hspace{-0.7cm} \frac{1}{N}\sum_{i\in U_{N}}E_{p}\left( L_{i\alpha
N1}^{2}\right) \\
&=&\frac{1}{N^{5}}E_{p}\left[ \left\{ \left(
\hat{t}_{y}-t_{y}\right) \left.
\frac{\partial \hat{m}_{i\alpha }^{*}}{\partial \left( N^{-1}\hat{t}%
_{i\alpha 4}\right) }\right| _{\hat{\mathbf{t}}_{i}=\mathbf{t}%
_{i}}\sum_{k\in U_{N}}z_{i\alpha k1}\left( \frac{I_{k}}{\pi
_{k}}-1\right)
\right\} ^{2}\right]  \\
&\leq &\frac{C}{N^{5}}\sum_{j,k,l,p\in U_{N}}\left| z_{i\alpha
j1}z_{i\alpha l1}y_{k}y_{p}\frac{\pi _{jl}-\pi _{j}\pi _{l}}{\pi
_{j}\pi _{l}}\frac{\pi _{kp}-\pi _{k}\pi _{p}}{\pi _{k}\pi
_{p}}\right| \leq \frac{C}{N^{3}}\sum_{k,p\in U_{N}}\left|
y_{k}y_{p}\right| \rightarrow 0.
\end{eqnarray*}
Next
\begin{eqnarray*}
\hspace{-0.35cm} && \hspace{-0.2cm} \frac{1}{N}\sum_{i\in U_{N}}E_{p}\left( L_{i\alpha N2}\right) ^{2} \\
\hspace{-0.35cm} &=& \hspace{-0.2cm} \frac{1}{N^{3}}E_{p}\left[ \left. \frac{\partial \hat{m}_{i\alpha }^{*}}{%
\partial \left( N^{-1}\hat{t}_{i\alpha 4}\right) }\right| _{\hat{\mathbf{t}}%
_{i\alpha}=\mathbf{t}_{i\alpha}}\sum_{k\in U_{N}}z_{i\alpha k1}\left(
\frac{I_{k}}{\pi _{k}}-1\right) \sum_{\beta \neq \alpha }\left\{
\hat{m}_{\beta }\left( x_{k\beta }\right) -\tilde{m}_{\beta
}\left( x_{k\beta }\right) \right\}
\right] ^{2} \\
\hspace{-0.35cm} &\leq& \hspace{-0.2cm} CN^{-3}\sum_{k\in U_{N}}\sum_{l\in U_{N}} E_{p}\left|
\left( \frac{I_{k}}{\pi _{k}}-1\right) \left( \frac{I_{l}}{\pi
_{l}}-1\right)
\right. \\
\hspace{-0.35cm} && \hspace{-0.2cm} \left. \times \sum_{\beta \neq \alpha }\sum_{\gamma \neq \alpha
}\left\{ \hat{m}_{\beta }\left( x_{k\beta }\right)
-\tilde{m}_{\beta }\left( x_{k\beta }\right) \right\} \left\{
\hat{m}_{\gamma }\left( x_{l\gamma }\right) -\tilde{m}_{\gamma
}\left( x_{l\gamma }\right) \right\} \right|.
\end{eqnarray*}
By Lemma \ref{LEM:sup(mtilde-hat)}, $%
N^{-1}\sum_{i\in U_{N}}E_{p}\left( L_{i\alpha N2}^{2}\right)
\rightarrow 0$, and the lemma follows immediately. \hfill
$\square$

\vspace{.2cm}
\noindent \textbf{A.4. Proofs of Theorems
\ref{THM:ADU}, \ref {THM:normality} and \ref{THM:GJLB}}

\begin{lemma}
\label{LEM:Epmtildestar-hatstar} Under Assumptions (A1)-(A8), for
the population and sample based SBLL estimators of ${m}\left(
x_{i\alpha
}\right) $ given in (\ref{DEF:mitildestar}) and (\ref{DEF:mihatstar}),
\[
\lim_{N\rightarrow \infty }\frac{1}{N}E_{p}\left[ \sum_{i\in
U_{N}}\left( \hat{m}_{i}^{*}-\tilde{m}_{i}^{*}\right) ^{2}\right]
=0.
\]
\end{lemma}

\noindent \textbf{Proof.} According to (\ref{EQ:Taylor
expansion2}), one has
\begin{eqnarray*}
&& \hspace{-0.3cm} \left. \frac{1}{N}E_{p}\left[ \sum_{i\in
U_{N}}\left(
\hat{m}_{i\alpha }^{*}-\tilde{m}_{i\alpha }^{*}\right) ^{2}\right] =\frac{1}{%
N^{3}}\sum_{i\in U_{N}}\sum_{k,l\in U_{N}}\Delta _{kl}\frac{z_{i\alpha k}}{%
\pi _{k}}\frac{z_{i\alpha l}}{\pi _{l}}\right. \\
&& \hspace{-0.3cm} -\frac{2}{N^{2}}\sum_{i,k\in U_{N}}z_{i\alpha
k}E_{p}\left[ \left( \frac{I_{k}}{\pi _{k}}-1\right) \left(
L_{i\alpha N}-R_{i\alpha N}\right) \right] +\frac{1}{N}\sum_{i\in
U_{N}}E_{p}\left( L_{i\alpha N}-R_{i\alpha N}\right) ^{2}.
\end{eqnarray*}
Following from Lemma 4 in \citep{BO00} and Assumption
(A7), the first term converges to zero as $N\rightarrow \infty $.
The third term also converges to zero by (\ref{EQ:remainder}) and
Lemma \ref{LEM:Remainders}. By the Cauchy-Schwartz
inequality, $\lim_{N\rightarrow \infty }\frac{1}{N}E_{p}\left[ \sum_{i\in
U_{N}}\left( \hat{m}_{i\alpha }^{*}-\tilde{m}_{i\alpha
}^{*}\right) ^{2}\right] =0$, $\alpha =1,...,d$. Note that
\begin{eqnarray*}
&&\sum_{i\in U_{N}}\left( \hat{m}_{i}^{*}-\tilde{m}_{i}^{*}\right)
^{2}=\sum_{i\in U_{N}}\left\{ \frac{1}{N}\left(
\hat{t}_{y}-t_{y}\right) +\sum_{\alpha =1}^{d}\left(
\hat{m}_{i\alpha }^{*}-\tilde{m}_{i\alpha
}^{*}\right) \right\} ^{2} \\
&=&\frac{1}{N}\left( \hat{t}_{y}-t_{y}\right) ^{2}+\frac{2}{N}\left( \hat{t}%
_{y}-t_{y}\right) \sum_{i\in U_{N}}\sum_{\alpha =1}^{d}\left( \hat{m}%
_{i\alpha }^{*}-\tilde{m}_{i\alpha }^{*}\right)+\sum_{i\in
U_{N}}\left\{ \sum_{\alpha =1}^{d}\left( \hat{m}_{i\alpha
}^{*}-\tilde{m}_{i\alpha }^{*}\right) \right\} ^{2}.
\end{eqnarray*}
By Assumption (A.7),
\[
\frac{1}{N^{2}}E_{p}\left( \hat{t}_{y}-t_{y}\right) ^{2} \leq \left( \frac{1}{\lambda }+\frac{n_{N}\max_{i,j\in
U_{N},i\neq j}\left| \Delta_{ij}\right| }{\lambda ^{2}}\right)
\frac{1}{N^{2}}\sum_{i\in U_{N}}y_{i}^{2}\rightarrow 0.
\]
Thus the desired result is obtained from the Cauchy-Schwartz inequality. \hfill
$\square$

\vskip .1in

\noindent \textbf{Proof of Theorem \ref{THM:ADU}.} Note that
$E_{p}\left[ I_{i}\right] =\pi _{i}$ and
\begin{equation}
\frac{\hat{t}_{y,{\mbox{\rm \tiny SBLL}}}-t_{y}}{N}=\frac{1}{N}\sum_{i\in U_{N}}%
(y_{i}-\tilde{m}_{i}^{*}) \left( I_{i}/\pi _{i}-1\right)
+\frac{1}{N}\sum_{i\in U_{N}}(\hat{m}_{i}^{*}-\tilde{m}_{i}^{*})\left( 1-
I_{i}/\pi _{i}\right) .  \label{EQ:tyhat-ty}
\end{equation}
Then
\begin{eqnarray}
E_{p}\left| \frac{\hat{t}_{y,{\mbox{\rm \tiny
SBLL}}}-t_{y}}{N}\right|
&\leq& \frac{1}{N} E_{p}\left| \sum_{i\in U_{N}}\left( y_{i}-\tilde{m}%
_{i}^{*}\right) \left( I_{i}/\pi _{i}-1\right) \right|  \label{EQ:ty_tilde-ty} \\
&&+\frac{1}{N^2}\left\{ E_{p}\left[ \sum_{i\in U_{N}}\left( \hat{m}_{i}^{*}-\tilde{m%
}_{i}^{*}\right) ^{2}\right] E_{p}\left[ \sum_{i\in
U_{N}}\left( 1-I_{i}/\pi _{i}\right)^{2}\right]
\right\} ^{1/2}. \nonumber
\end{eqnarray}
According to Assumptions (A1)-(A6), $\limsup_{N\rightarrow \infty }\frac{1}{N%
}\sum_{i\in U_{N}}\left( y_{i}-\tilde{m}_{i}^{*}\right)
^{2}<\infty$. Following the same arguments of Theorem 1 in \citep{BO00}, the
first term on the right of (\ref{EQ:ty_tilde-ty}) converges to zero as $%
N\rightarrow \infty $. For the second term, (A7) implies that
\[
E_{p}\left[ \frac{1}{N}\sum_{i\in U_{N}}\left( 1-I_{i}/\pi _{i}\right) ^{2}
\right] =\sum_{i\in U_{N}}\frac{\pi _{i}\left( 1-\pi _{i}\right)
}{N\pi _{i}^{2}}\leq \frac{1}{\lambda }.
\]
According to Lemma \ref{LEM:Epmtildestar-hatstar},
$\lim_{N\rightarrow
\infty } \frac{1}{N}\sum_{i\in U_{N}}E_{p}\left[ \left( \hat{m}_{i}^{*}-%
\tilde{m}_{i}^{*}\right) ^{2}\right] \rightarrow 0$ and the result follows from the Markov's
inequality. \hfill $\square$

The next theorem is to derive the asymptotic mean squared error of
the proposed spline estimator in (\ref{DEF:tySBLLhat}).

\begin{theorem}
\label{THM:variance} Under Assumptions (A1)-(A8),
\begin{equation}
n_{N}E_{p}\left( \frac{\hat{t}_{y,\SBLL}-t_{y}}{N}\right)
^{2}=\frac{n_{N}}{N^{2}}\sum_{i,j\in U_{N}}\Delta _{ij}\frac{y_{i}-\tilde{m}%
_{i}^{*}}{\pi _{i}}\frac{y_{j}-\tilde{m}_{j}^{*}}{\pi
_{j}}+o\left( 1\right) .  \label{EQ:AMSE}
\end{equation}
\end{theorem}

Denote
\begin{equation}
\text{AMSE}\left( N^{-1}\hat{t}_{y,{\mbox{\rm \tiny SBLL}}}\right) =\frac{1}{%
N^{2}}\sum_{i,j\in U_{N}}\Delta_{ij}\frac{
y_{i}-\tilde{m}_{i}^{*}}{\pi _{i}} \frac{
y_{j}-\tilde{m}_{j}^{*}}{\pi _{j}}\label{DEF:AMSE}
\end{equation}
the asymptotic mean squared error in (\ref{EQ:AMSE}). The next
result shows that it can be estimated consistently by
$\widehat{V}\left( N^{-1}\hat{t} _{y,{\mbox{\rm \tiny
SBLL}}}\right) $ in (\ref{DEF:Vhat}).

\begin{theorem}
\label{THM:Vhat-AMSE} Under (A1)-(A8),
\[
\lim_{N\rightarrow \infty
}n_{N}E_{p}\left| \widehat{V}\left( N^{-1}\hat{t}_{y,{\mbox{\rm \tiny SBLL}}%
}\right) -\mathrm{{AMSE}\left( N^{-1}\hat{t}_{y,{\mbox{\rm \tiny SBLL}}%
}\right) }\right| =0.
\]
\end{theorem}

The proofs of Theorems \ref{THM:variance} and \ref{THM:Vhat-AMSE} are somewhat
trivial and we refer the readers to \citep{WW11}.

\vskip .1in

\noindent \textbf{Proof of Theorem \ref{THM:normality}.} According
to (\ref {EQ:tyhat-ty}),
\begin{eqnarray*}
\frac{\hat{t}_{y,{\mbox{\rm \tiny SBLL}}}-t_{y}}{N}=\frac{\tilde{t}_{y,{%
\mbox{\rm \tiny SBLL}}}-t_{y}}{N}+\sum_{i \in U_{N}} \frac{\tilde{m}_{i}^{*}-%
\hat{m}_{i}^{*}}{N}\left(\frac{I_{i}}{\pi_{i}}-1\right).
\end{eqnarray*}
From the proof of Theorem \ref{THM:variance}, $\sum_{i \in U_{N}} \frac{%
\tilde{m}_{i}^{*}-\hat{m}_{i}^{*}}{N}\left(\frac{I_{i}}{\pi_{i}}%
-1\right)=o_{p}\left( n_{N}^{-1/2}\right).$ Theorem
\ref{THM:Vhat-AMSE}
implies that $\widehat{V}\left( N^{-1}\hat{t}_{y,{\mbox{\rm \tiny SBLL}}%
}\right)/\text{AMSE}\left( N^{-1} \hat{t}_{y,{\mbox{\rm \tiny
SBLL}}}\right) \rightarrow 1$ in probability. The desired result
follows. \hfill $\square $

\noindent \textbf{Proof of Theorem \ref{THM:GJLB}.} Let
\begin{eqnarray*}
T_{1} &=&\frac{n_{N}^{1/2}}{N}\sum_{i\in U_{N}}\left\{ \tilde{m}%
_{i}^{*}-m\left( \mathbf{x}_{i}\right) \right\} \left(
1-\frac{I_{i}}{\pi
_{i}}\right) ,\ T_{2}=\frac{n_{N}^{1/2}}{N}\sum_{i\in U_{N}}\left( \hat{m%
}_{i}^{*}-\tilde{m}_{i}^{*}\right) \left( 1-\frac{I_{i}}{\pi
_{i}}\right) ,
\\
T_{3} &=&\frac{n_{N}^{1/2}}{N}\sum_{i\in U_{N}}\sigma \left( \mathbf{x}%
_{i}\right) \varepsilon _{i}\left( \frac{I_{i}}{\pi _{i}}-1\right)
.
\end{eqnarray*}
Then $n_{N}^{1/2}N^{-1}\left( \hat{t}_{y,{\mbox{\rm \tiny SBLL}}%
}-t_{y}\right) $ can be represented as the sum of $T_{1}$, $T_{2}$ and $%
T_{3} $. For the first term,
\begin{eqnarray*}
ET_{1}^{2} &=&\frac{n_{N}}{N^{2}}\sum_{i,j\in U_{N}}\left[ E\left(
m\left(
\mathbf{x}_{i}\right) -\tilde{m}_{i}^{*}\right) \left( m\left( \mathbf{x}%
_{j}\right) -\tilde{m}_{j}^{*}\right) \right] \frac{\Delta
_{ij}}{\pi
_{i}\pi _{j}} \\
&\leq &\frac{n_{N}}{N}\left( \frac{1}{\lambda
}+\frac{N\max_{i,j\in
U_{N},i\neq j}\left| \Delta _{ij}\right| }{\lambda ^{2}}\right) \frac{1}{N}%
\sum_{i\in U_{N}}E\left\{ m\left( \mathbf{x}_{i}\right) -\tilde{m}%
_{i}^{*}\right\} ^{2}.
\end{eqnarray*}
By Theorem 2.1 in \citep{WY07}, $\left| m\left(
\mathbf{x}_{i}\right) -\tilde{m}_{i}^{*}\right| =o_{p}\left(
n^{-2/5}\log n\right) $, for any $i\in U_{N}$, which implies that
$ET_{1}^{2}\rightarrow 0$. Now
for $T_{2}$%
\[
ET_{2}^{2}\leq \frac{n_{N}}{N}\left( \frac{1}{\lambda
}+\frac{N\max_{i,j\in
U_{N},i\neq j}\left| \Delta _{ij}\right| }{\lambda ^{2}}\right) \frac{1}{N}%
\sum_{i\in U_{N}}E\left( \hat{m}_{i}^{*}-\tilde{m}_{i}^{*}\right)
^{2}.
\]
By Lemma \ref{LEM:sup(mtilde-hat)}, $ET_{2}^{2}\rightarrow 0$.
Finally,
\[
ET_{3}^{2}=\frac{n_{N}}{N^{2}}\sum_{i\in U_{N}}\sigma
^{2}\left(
\mathbf{x}_{i}\right) \frac{1-\pi _{i}}{\pi _{i}}\leq \frac{n_{N}}{N\lambda }%
\frac{1}{N}\sum_{i\in U_{N}}\sigma ^{2}\left(
\mathbf{x}_{i}\right) ,
\]
\[
\limsup_{N\rightarrow \infty }ET_{3}^{2}\leq \frac{1}{\lambda }%
\limsup_{N\rightarrow \infty }\frac{1}{N}\sum_{i\in U_{N}}\sigma
^{2}\left( \mathbf{x}_{i}\right) <\infty .
\]
By the Cauchy-Schwartz inequality the cross product terms go to zero as $%
N\rightarrow \infty $. The desired result follows. \hfill
$\square$

\section*{Acknowledgment}

The research work of the first author was supported by NSF grant DMS-0905730.
The authors thank Professor Lijian Yang for helpful discussions. The authors
would also like to thank two anonymous referees for their insightful comments.

\end{document}